\documentclass[12pt,twocolumn]{aastex631}
\usepackage{hyperref}
\usepackage{color}
\usepackage{graphicx}
\usepackage{xfrac}
\usepackage{ulem}
\usepackage{xcolor}
\usepackage{amsmath}
\usepackage{afterpage}
\usepackage{longtable}
\newcommand{\tam}{\color{black}}

\newcommand{\oii}{[\ion{O}{2}]$\lambda\lambda 3726,3729$ }

\shorttitle{DESI Galaxy Visual Inspection}
\shortauthors{Lan et al.}

\begin{document}
\title{The DESI Survey Validation: Results from Visual Inspection of Bright Galaxies, Luminous Red Galaxies, and Emission Line Galaxies} 
\correspondingauthor{Ting-Wen Lan}
\email{twlan@ntu.edu.tw}
\author{Ting-Wen Lan}
\affiliation{Graduate Institute of Astrophysics and Department of Physics, National Taiwan University, No. 1, Sec. 4, Roosevelt Rd., Taipei 10617, Taiwan}
\affiliation{Department of Astronomy and Astrophysics, University of California, Santa Cruz, 1156 High Street, Santa Cruz, CA 95065, USA}
\author{R.~Tojeiro}
\affiliation{SUPA, School of Physics and Astronomy, University of St Andrews, St Andrews, KY16 9SS, UK}
\author{E.~Armengaud}
\affiliation{IRFU, CEA, Universit\'{e} Paris-Saclay, F-91191 Gif-sur-Yvette, France}
\author{J.~Xavier~Prochaska}
\affiliation{Department of Astronomy and Astrophysics, University of California, Santa Cruz, 1156 High Street, Santa Cruz, CA 95065, USA}
\affiliation{Kavli Institute for the Physics and Mathematics of the Universe (Kavli IPMU), WPI, The University of Tokyo Institutes for Advanced Study (UTIAS), The
University of Tokyo, Kashiwa, Chiba, Kashiwa 277-8583, Japan}
\author{T.~M.~Davis}
\affiliation{School of Mathematics and Physics, University of Queensland, 4072, Australia}
\author{David~M.~Alexander}
\affiliation{Centre for Extragalactic Astronomy, Department of Physics, Durham University, South Road, Durham, DH1 3LE, UK}
\author{A.~Raichoor}
\affiliation{Lawrence Berkeley National Laboratory, 1 Cyclotron Road, Berkeley, CA 94720, USA}
\author{Rongpu Zhou}
\affiliation{Lawrence Berkeley National Laboratory, 1 Cyclotron Road, Berkeley, CA 94720, USA}
\author{Christophe~Yèche}
\affiliation{IRFU, CEA, Universit\'{e} Paris-Saclay, F-91191 Gif-sur-Yvette, France}
\author{C.~Balland}
\affiliation{Sorbonne Universit\'{e}, CNRS/IN2P3, Laboratoire de Physique Nucl\'{e}aire et de Hautes Energies (LPNHE), FR-75005 Paris, France}
\author{S.~BenZvi}
\affiliation{Department of Physics \& Astronomy, University of Rochester, 206 Bausch and Lomb Hall, P.O. Box 270171, Rochester, NY 14627-0171, USA}
\author{A.~Berti}
\affiliation{Department of Physics and Astronomy, The University of Utah, 115 South 1400 East, Salt Lake City, UT 84112, USA}
\author{R.~Canning}
\affiliation{Institute of Cosmology \& Gravitation, University of Portsmouth, Dennis Sciama Building, Portsmouth, PO1 3FX, UK}
\author{A.~Carr}
\affiliation{School of Mathematics and Physics, University of Queensland, 4072, Australia}
\author{H.~Chittenden}
\affiliation{SUPA, School of Physics and Astronomy, University of St Andrews, St Andrews, KY16 9SS, UK}
\author{S.~Cole}
\affiliation{Institute for Computational Cosmology, Department of Physics, Durham University, South Road, Durham DH1 3LE, UK}
\author{M.-C.~Cousinou}
\affiliation{Aix Marseille Univ, CNRS/IN2P3, CPPM, Marseille, France}
\author{K.~Dawson}
\affiliation{Department of Physics and Astronomy, The University of Utah, 115 South 1400 East, Salt Lake City, UT 84112, USA}
\author{Biprateep~Dey}
\affiliation{Department of Physics \& Astronomy and Pittsburgh Particle Physics, Astrophysics, and Cosmology Center (PITT PACC), University of Pittsburgh, 3941 O'Hara Street, Pittsburgh, PA 15260, USA}
\author{K.~Douglass}
\affiliation{Department of Physics \& Astronomy, University of Rochester, 206 Bausch and Lomb Hall, P.O. Box 270171, Rochester, NY 14627-0171, USA}
\author{A.~Edge}
\affiliation{Institute for Computational Cosmology, Department of Physics, Durham University, South Road, Durham DH1 3LE, UK}
\author{S.~Escoffier}
\affiliation{Aix Marseille Univ, CNRS/IN2P3, CPPM, Marseille, France}
\author{A.~Glanville}
\affiliation{School of Mathematics and Physics, University of Queensland, 4072, Australia}
\author{S.~Gontcho A Gontcho}
\affiliation{Lawrence Berkeley National Laboratory, 1 Cyclotron Road, Berkeley, CA 94720, USA}
\author{J.~Guy}
\affiliation{Lawrence Berkeley National Laboratory, 1 Cyclotron Road, Berkeley, CA 94720, USA}
\author{C.~Hahn}
\affil{Department of Astrophysical Sciences, Princeton University, Peyton Hall, Princeton NJ 08544, USA}
\affil{Lawrence Berkeley National Laboratory, One Cyclotron Road, Berkeley CA 94720, USA}
\author{C.~Howlett}
\affiliation{School of Mathematics and Physics, University of Queensland, 4072, Australia}

\author{
Ho Seong Hwang}
\affiliation{Astronomy Program, Department of Physics and Astronomy, Seoul National University, 1 Gwanak-ro, Gwanak-gu, Seoul 08826, Republic of Korea}
\affiliation{SNU Astronomy Research Center, Seoul National University, 1 Gwanak-ro, Gwanak-gu, Seoul 08826, Republic of Korea}
\affiliation{Korea Astronomy and Space Science Institute (KASI), 776 Daedeokdae-ro, Yuseong-gu, Daejeon 34055, Republic of Korea}

\author{L.~Jiang}
\affiliation{Kavli Institute for Astronomy and Astrophysics at Peking University, PKU, 5 Yiheyuan Road, Haidian District, Beijing 100871, P.R. China}
\author{A.~Kov\'acs}
\affiliation{Departamento de Astrof\'{\i}sica, Universidad de La Laguna (ULL), E-38206, La Laguna, Tenerife, Spain}
\affiliation{Instituto de Astrof\'{i}sica de Canarias, C/ Vía L\'{a}ctea, s/n, 38205 San Crist\'{o}bal de La Laguna, Santa Cruz de Tenerife, Spain}
\author{M.~Mezcua}
\affiliation{Institut d'Estudis Espacials de Catalunya (IEEC), 08034 Barcelona, Spain}
\affiliation{Institute of Space Sciences, ICE-CSIC, Campus UAB, Carrer de Can Magrans s/n, 08913 Bellaterra, Barcelona, Spain}
\author{S.~Moore}
\affiliation{Institute for Computational Cosmology, Department of Physics, Durham University, South Road, Durham DH1 3LE, UK}
\author{S.~Nadathur}
\affiliation{Department of Physics \& Astronomy, University College London, Gower Street, London, WC1E 6BT, UK}
\affiliation{Institute of Cosmology \& Gravitation, University of Portsmouth, Dennis Sciama Building, Portsmouth, PO1 3FX, UK}
\author{M.~Oh}
\affiliation{Department of Astronomy, Shanghai Jiao Tong University, Shanghai 200240, China}
\author{D.~Parkinson}
\affiliation{Korea Astronomy and Space Science Institute, 776, Daedeokdae-ro, Yuseong-gu, Daejeon 34055, Republic of Korea}

\author{A.~Rocher}
\affiliation{IRFU, CEA, Universit\'{e} Paris-Saclay, F-91191 Gif-sur-Yvette, France}
\author{A.~J.~Ross}
\affiliation{Center for Cosmology and AstroParticle Physics, The Ohio State University, 191 West Woodruff Avenue, Columbus, OH 43210, USA}
\author{V.~Ruhlmann-Kleider}
\affiliation{IRFU, CEA, Universit\'{e} Paris-Saclay, F-91191 Gif-sur-Yvette, France}
\author{C.~G.~Sabiu}
\affiliation{Natural Science Research Institute, University of Seoul, 163 Seoulsiripdae-ro, Dongdaemun-gu, Seoul, South Korea}
\author{K.~Said}
\affiliation{School of Mathematics and Physics, University of Queensland, 4072, Australia}
\author{C.~Saulder}
\affiliation{Korea Astronomy and Space Science Institute, 776, Daedeokdae-ro, Yuseong-gu, Daejeon 34055, Republic of Korea}
\author{D.~Sierra-Porta}
\affiliation{Universidad Tecnológica de Bolivar. Facultad de Ciencias Básicas. Cartagena de Indias 130010, Colombia}
\affiliation{Departamento de F\'isica, Universidad de los Andes, Cra. 1 No. 18A-10, Edificio Ip, CP 111711, Bogot\'a, Colombia}
\author{B.~Weiner}
\affiliation{Steward Observatory, University of Arizona, 933 N, Cherry Ave, Tucson, AZ 85721, USA}
\author{J.~Yu}
\affiliation{Ecole Polytechnique F\'{e}d\'{e}rale de Lausanne, CH-1015 Lausanne, Switzerland}
\author{P.~Zarrouk}
\affiliation{Sorbonne Universit\'{e}, CNRS/IN2P3, Laboratoire de Physique Nucl\'{e}aire et de Hautes Energies (LPNHE), FR-75005 Paris, France}
\author{Y.~Zhang}
\affiliation{Department of Physics and Center for Cosmology and Particle Physics, New York University, New York, NY 10003, USA}
\author{H.~Zou}
\affiliation{National Astronomical Observatories, Chinese Academy of Sciences, A20 Datun Rd., Chaoyang District, Beijing, 100012, P.R. China}
\author{S.~Ahlen}
\affiliation{Physics Dept., Boston University, 590 Commonwealth Avenue, Boston, MA 02215, USA}
\author{S.~Bailey}
\affiliation{Lawrence Berkeley National Laboratory, 1 Cyclotron Road, Berkeley, CA 94720, USA}
\author{D.~Brooks}
\affiliation{Department of Physics \& Astronomy, University College London, Gower Street, London, WC1E 6BT, UK}
\author{A.P.~Cooper}
\affiliation{Institute of Astronomy and Department of Physics, National Tsing Hua University, 101 Kuang-Fu Rd. Sec. 2, Hsinchu 30013, Taiwan}
\author{A.~de la Macorra}
\affiliation{Instituto de F\'{\i}sica, Universidad Nacional Aut\'{o}noma de M\'{e}xico,  Cd. de M\'{e}xico  C.P. 04510,  M\'{e}xico}
\author{A.~Dey}
\affiliation{NSF's National Optical-Infrared Astronomy Research Laboratory, 950 N. Cherry Avenue, Tucson, AZ 85719, USA}
\author{G.~Dhungana}
\affiliation{Department of Physics, Southern Methodist University, 3215 Daniel Avenue, Dallas, TX 75275, USA}
\author{P.~Doel}
\affiliation{Department of Physics \& Astronomy, University College London, Gower Street, London, WC1E 6BT, UK}
\author{S.~Eftekharzadeh}
\affiliation{Universities Space Research Association, NASA Ames Research Centre}
\author{K.~Fanning}
\affiliation{Department of Physics, University of Michigan, Ann Arbor, MI 48109, USA}
\author{A.~Font-Ribera}
\affiliation{Institut de F\'{i}sica d’Altes Energies (IFAE), The Barcelona Institute of Science and Technology, Campus UAB, 08193 Bellaterra Barcelona, Spain}
\author{L.~Garrison}
\affiliation{Flatiron Institute, 162 5th Avenue, New York, NY 10010, USA}
\author{E.~Gaztañaga}
\affiliation{Institut d'Estudis Espacials de Catalunya (IEEC), 08034 Barcelona, Spain}
\affiliation{Institute of Space Sciences, ICE-CSIC, Campus UAB, Carrer de Can Magrans s/n, 08913 Bellaterra, Barcelona, Spain}
\author{R.~Kehoe}
\affiliation{Department of Physics, Southern Methodist University, 3215 Daniel Avenue, Dallas, TX 75275, USA}
\author{T.~Kisner}
\affiliation{Lawrence Berkeley National Laboratory, 1 Cyclotron Road, Berkeley, CA 94720, USA}
\author{A.~Kremin}
\affiliation{Department of Physics, University of Michigan, Ann Arbor, MI 48109, USA}
\affiliation{Lawrence Berkeley National Laboratory, 1 Cyclotron Road, Berkeley, CA 94720, USA}
\affiliation{Physics Department, University of Michigan Ann Arbor, MI 48109, USA}
\author{M.~Landriau}
\affiliation{Lawrence Berkeley National Laboratory, 1 Cyclotron Road, Berkeley, CA 94720, USA}
\author{L.~Le~Guillou}
\affiliation{Sorbonne Universit\'{e}, CNRS/IN2P3, Laboratoire de Physique Nucl\'{e}aire et de Hautes Energies (LPNHE), FR-75005 Paris, France}
\author{Michael E.~Levi}
\affiliation{Lawrence Berkeley National Laboratory, 1 Cyclotron Road, Berkeley, CA 94720, USA}
\author{C.~Magneville}
\affiliation{IRFU, CEA, Universit\'{e} Paris-Saclay, F-91191 Gif-sur-Yvette, France}
\author{Aaron M. Meisner}
\affiliation{NSF's National Optical-Infrared Astronomy Research Laboratory, 950 N. Cherry Avenue, Tucson, AZ 85719, USA}
\author{R.~Miquel}
\affiliation{Instituci\'{o} Catalana de Recerca i Estudis Avan\c{c}ats, Passeig de Llu\'{\i}s Companys, 23, 08010 Barcelona, Spain}
\affiliation{Institut de F\'{i}sica d’Altes Energies (IFAE), The Barcelona Institute of Science and Technology, Campus UAB, 08193 Bellaterra Barcelona, Spain}
\author{J.~Moustakas}
\affiliation{Department of Physics and Astronomy, Siena College, 515 Loudon Road, Loudonville, NY 12211, USA}
\author{Adam~D.~Myers}
\affiliation{Department of Physics \& Astronomy, University  of Wyoming, 1000 E. University, Dept.~3905, Laramie, WY 82071, USA}
\author{Jeffrey A.~Newman}
\affiliation{Department of Physics \& Astronomy and Pittsburgh Particle Physics, Astrophysics, and Cosmology Center (PITT PACC), University of Pittsburgh, 3941 O'Hara Street, Pittsburgh, PA 15260, USA}
\author{J.D.Nie}
\affiliation{National Astronomical Observatories, Chinese Academy of Sciences, A20 Datun Rd., Chaoyang District, Beijing, 100012, P.R. China}
\author{N.~Palanque-Delabrouille}
\affiliation{IRFU, CEA, Universit\'{e} Paris-Saclay, F-91191 Gif-sur-Yvette, France}
\affiliation{Lawrence Berkeley National Laboratory, 1 Cyclotron Road, Berkeley, CA 94720, USA}
\author{W.J.~Percival}
\affiliation{Department of Physics and Astronomy, University of Waterloo, 200 University Ave W, Waterloo, ON N2L 3G1, Canada}
\affiliation{Perimeter Institute for Theoretical Physics, 31 Caroline St. North, Waterloo, ON N2L 2Y5, Canada}
\affiliation{Waterloo Centre for Astrophysics, University of Waterloo, 200 University Ave W, Waterloo, ON N2L 3G1, Canada}
\author{C.~Poppett}
\affiliation{Lawrence Berkeley National Laboratory, 1 Cyclotron Road, Berkeley, CA 94720, USA}
\affiliation{Space Sciences Laboratory, University of California, Berkeley, 7 Gauss Way, Berkeley, CA  94720, USA}
\affiliation{University of California, Berkeley, 110 Sproul Hall \#5800 Berkeley, CA 94720, USA}
\author{F.~Prada}
\affiliation{Instituto de Astrofisica de Andaluc\'{i}a, Glorieta de la Astronom\'{i}a, s/n, E-18008 Granada, Spain}
\author{M.~Schubnell}
\affiliation{Department of Physics, University of Michigan, Ann Arbor, MI 48109, USA}
\author{Gregory~Tarl\'{e}}
\affiliation{Department of Physics, University of Michigan, Ann Arbor, MI 48109, USA}
\author{B.~A.~Weaver}
\affiliation{NSF's National Optical-Infrared Astronomy Research Laboratory, 950 N. Cherry Avenue, Tucson, AZ 85719, USA}
\author{K.~Zhang}
\affiliation{Lawrence Berkeley National Laboratory, 1 Cyclotron Road, Berkeley, CA 94720, USA}
\author{Zhimin~Zhou}
\affiliation{National Astronomical Observatories, Chinese Academy of Sciences, A20 Datun Rd., Chaoyang District, Beijing, 100012, P.R. China}
\begin{abstract}
The Dark Energy Spectroscopic Instrument (DESI) Survey has obtained a set of spectroscopic measurements of galaxies to validate the final survey design and target selections. To assist in these tasks, we visually inspect (VI) DESI spectra of approximately 2,500 bright galaxies, 3,500 luminous red galaxies (LRGs), and 10,000 emission line galaxies (ELGs), to obtain robust redshift identifications. We then utilize the VI redshift information to characterize the performance of the DESI operation. Based on the VI catalogs, our results show that the final survey design yields samples of bright galaxies, LRGs, and ELGs with purity greater than $99\%$. Moreover, we demonstrate that the precision of the redshift measurements is approximately 10 km/s for bright galaxies and ELGs and approximately 40 km/s for LRGs. The average redshift accuracy is within 10 km/s for the three types of galaxies. The VI process also helps improve the quality of the DESI data by identifying spurious spectral features introduced by the pipeline. Finally, we show examples of unexpected real astronomical objects, such as Ly$\alpha$ emitters and strong lensing candidates, identified by VI. These results demonstrate the importance and utility of visually inspecting data from incoming and upcoming surveys, especially during their early operation phases. 
\end{abstract}
\section{Introduction}
{\tam The Dark Energy Spectroscopic Instrument \citep[DESI,][]{Levi2013} survey is a stage-IV experiment for probing the nature of dark energy. It will obtain tens of millions of spectra of stars, galaxies, and quasars and use the 3D positions of extragalactic sources to detect the signals of baryon acoustic oscillations \citep[e.g.,][]{Eisenstein2005} as a function of redshift. By doing so, DESI will measure the expansion rate of the Universe across cosmic time and constrain the equation of state of dark energy. In addition, DESI will probe the growth of cosmic structure via redshift-space distortion and constrain the summed neutrino masses. All the measurements and constraints will be obtained with unprecedented precision \citep{DESI2016, DESI2016b}.}

To achieve its scientific goals, DESI has been designed to observe four types of extragalactic sources: 
\begin{itemize}
\item \textbf{Bright galaxies (bright galaxy survey, BGS)} \citep{BGSTS,BGSp}, a galaxy sample whose redshifts can be obtained during the bright time of the observations together with the Milky Way stars \citep{MWTS,MWSp}. {\tam The BGS consists of two selections, (1) BGS bright --- a magnitude limit sample with $r<19.5$, which is similar to the depth of the Galaxy And Mass Assembly (GAMA) survey \citep{gama} and about 2 magnitude deeper than the main galaxy sample from the Sloan Digital Sky Survey \citep{SDSSMGS}, and (2) BGS faint --- a sample of fainter $19.5 < r < 20.175$ galaxies with a color-selection to achieve high redshift efficiency.} 
The {\tam designed } redshift coverage of BGS is from $z\sim0$ to $z\sim0.5$.

\item \textbf{Luminous red galaxies (LRGs)} \citep{LRGTS,LRGp}, a population of massive passive galaxies selected based on their colors and brightness. {\tam The DESI LRG sample has a significantly higher target density than that of any previous LRG surveys, including the SDSS LRG survey \citep{eisenstein2001}, Baryon Oscillation Spectroscopic Survey (BOSS) \citep{reid2016} and extended Baryon Oscillation Spectroscopic Survey (eBOSS) \citep{prakash2016}, and also probes higher redshifts with a designed redshift coverage from $z\sim0.4$ to $z\sim1.1$.}

\item \textbf{Emission line galaxies (ELGs)} \citep{ELGTS,ELGp}, a color-magnitude selected population of star-forming galaxies that emit \oii strong lines.  The {\tam designed} redshift coverage of ELGs is from $z\sim0.6$ to $z\sim1.6$. {\tam This emission-line galaxy population has been revealed by deep spectroscopic surveys across a few square degrees, e.g. DEEP2 survey \citep{Newman2013} and has been used as a tracer of the large-scale structure, e.g. the WiggleZ Dark Energy Survey \citep{Blake2011} and eBOSS ELG survey \citep{Raichoor2021}. The DESI ELG survey will offer a sample for cosmological measurements with 10 times higher target density by observing galaxies up
to one magnitude fainter than the eBOSS ELG sample \citep{ELGTS}.}

\item \textbf{Quasars (QSOs)} \citep{QSOTS,QSOp}, {\tam a quasar population selected by a Random
Forests algorithm based on the color and magnitude information of the sources. The DESI QSO sample is designed to cover a wide range of redshifts from the local Universe to $z\sim3$. Quasars with $z<2$ will be used primarily as tracers of the matter distribution, while quasars with $z>2$ will be used primarily as background light to probe the matter distribution traced by the Lyman $\alpha$ forest in the foreground. DESI aims to obtain approximately three million quasar spectra and quadruple the number of known quasars obtained from previous surveys \citep[e.g. eBOSS,][]{Dawson2016,Lyke2020}.}

\end{itemize}
Together, these sources will enable DESI to map the matter density distribution of the Universe across 10 billion years of cosmic time from the local Universe to $z\sim2.5$. 
However, to meet the requirements of the DESI design, the sources are selected from a novel parameter space. For example, {\tam DESI is the first large-scale survey to observe a high density of ELGs with $1.1 < z < 1.6$}. {\tam Similarly, our BGS and LRG surveys will cover fainter objects that have not been extensively observed in previous surveys.} 

In order to maximize the efficiency of obtaining redshifts of extragalactic sources covering the desired redshift ranges, a set of Survey Validation (SV) observations \citep{SVoverview} were designed to cover a wider range of parameter space with longer exposure times than the Main Survey. This SV dataset is used to explore and define the final target selections for the DESI Main Survey.

To support such analyses, one of the key components of the DESI SV is to construct catalogs of the DESI targets with robust redshift identifications. While an automatic data pipeline \citep{pipeline, Redrock} has been developed to reduce the raw data and obtain the best-fit redshifts of spectra, it is crucial to have an additional validation of the dataset which enables us to (1) assess the quality of data; (2) assess the precision and accuracy of the redshift measurements from the pipeline; and (3) offer feedback for improving the performance of the data pipeline, especially during the early phase of the survey. 
To this end, we have visually inspected (VI) more than 15,000 DESI SV spectra and produced VI catalogs of BGS, LRGs, ELGs and QSOs observed in the DESI SV program. Visual inspection of survey datasets has been demonstrated to be essential during the development of a survey \citep[e.g.,][]{Newman2013,Comparat2016,Paris2018}.

In this work, we describe the procedure that we adopt for visually inspecting DESI spectra. We make use of the VI catalogs to explore the target selections, and to validate the design of the DESI Main Survey. 
Given that galaxy and quasar spectra have different properties, in this work we focus on the VI catalogs of galaxies, including BGS, LRGs and ELGs. The QSO VI catalog and results are described in \citet{QSOVI}. 

The structure of the paper is as follows. The procedure of VI is described in Section 2. The data and the VI catalogs are described in Section 3. 
We show our analysis for validating the DESI design in Section 4 and demonstrate how this VI process assists the development of the DESI pipeline in Section 5. In Section 6, we show some spectra of rare galaxy types identified during the VI process and highlight scientific projects that can be done with the incoming DESI datasets. We summarize in Section 7. 

\begin{figure*}
\center
\includegraphics[width=0.8\textwidth]{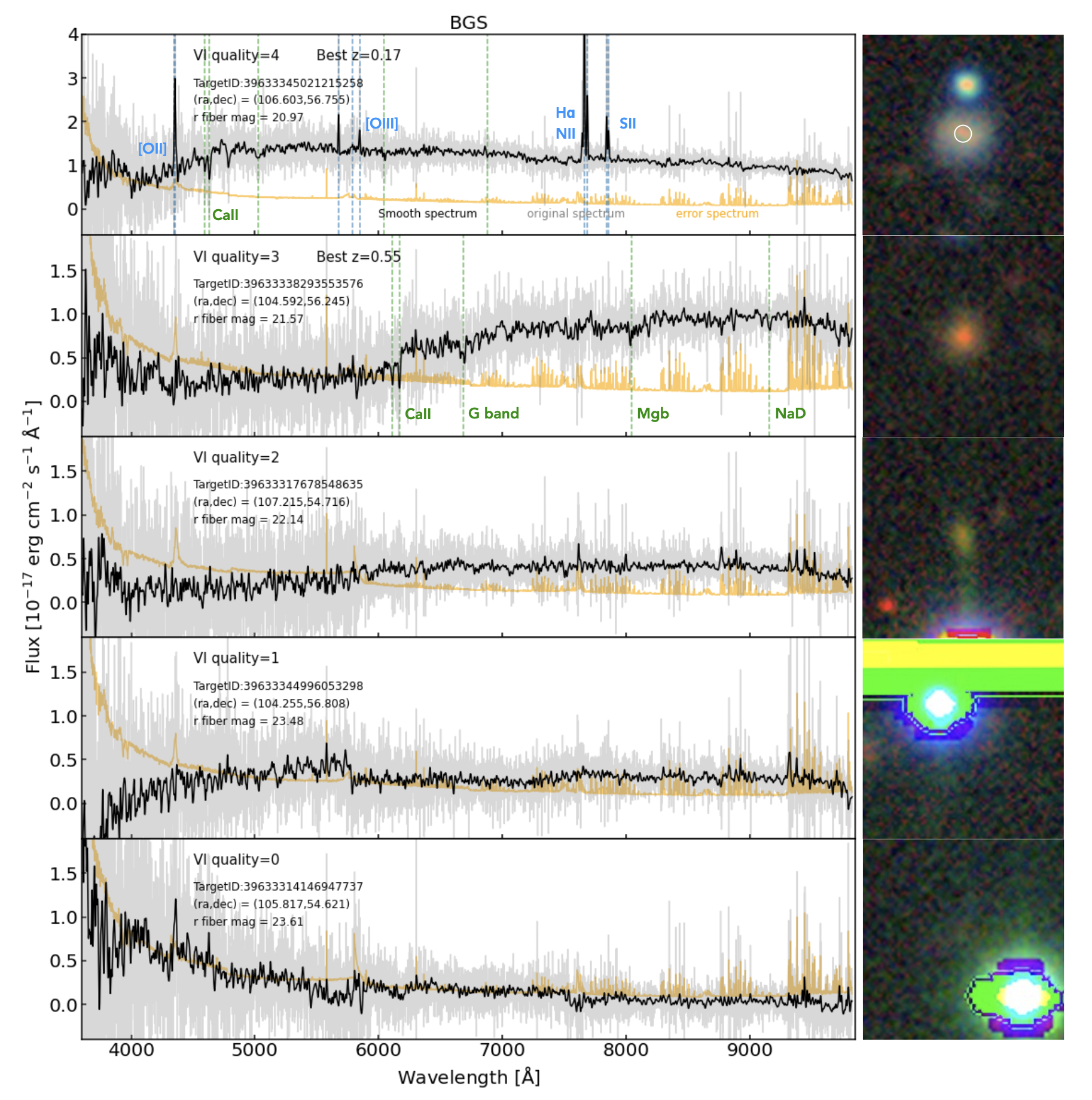}
\caption{{\tam Example of BGS spectra, ordered by their VI quality values} from top with VI quality 4 (best quality)
to bottom with VI quality 0 (poorest quality). The quality 4 spectrum has multiple spectral features including both emission and absorption lines. The quality 3 spectrum shows strong Ca II absorption lines with weak NaD and Mgb absorption lines. Spectra with quality 2 and below do not have spectral features that can be robustly identifiable. The right panels show the DESI Legacy Survey images of the BGS targets. The spectra in grey, black and orange colors are the original observed galaxy spectrum, the smooth spectrum with a Gaussian filter, and the error spectrum, respectively. The side length of the image is 18" and the white circle in the top panel image reflects the DESI fiber size.} 
\label{fig:BGS_spectra}
\end{figure*}

\begin{figure*}
\center
\includegraphics[width=0.8\textwidth]{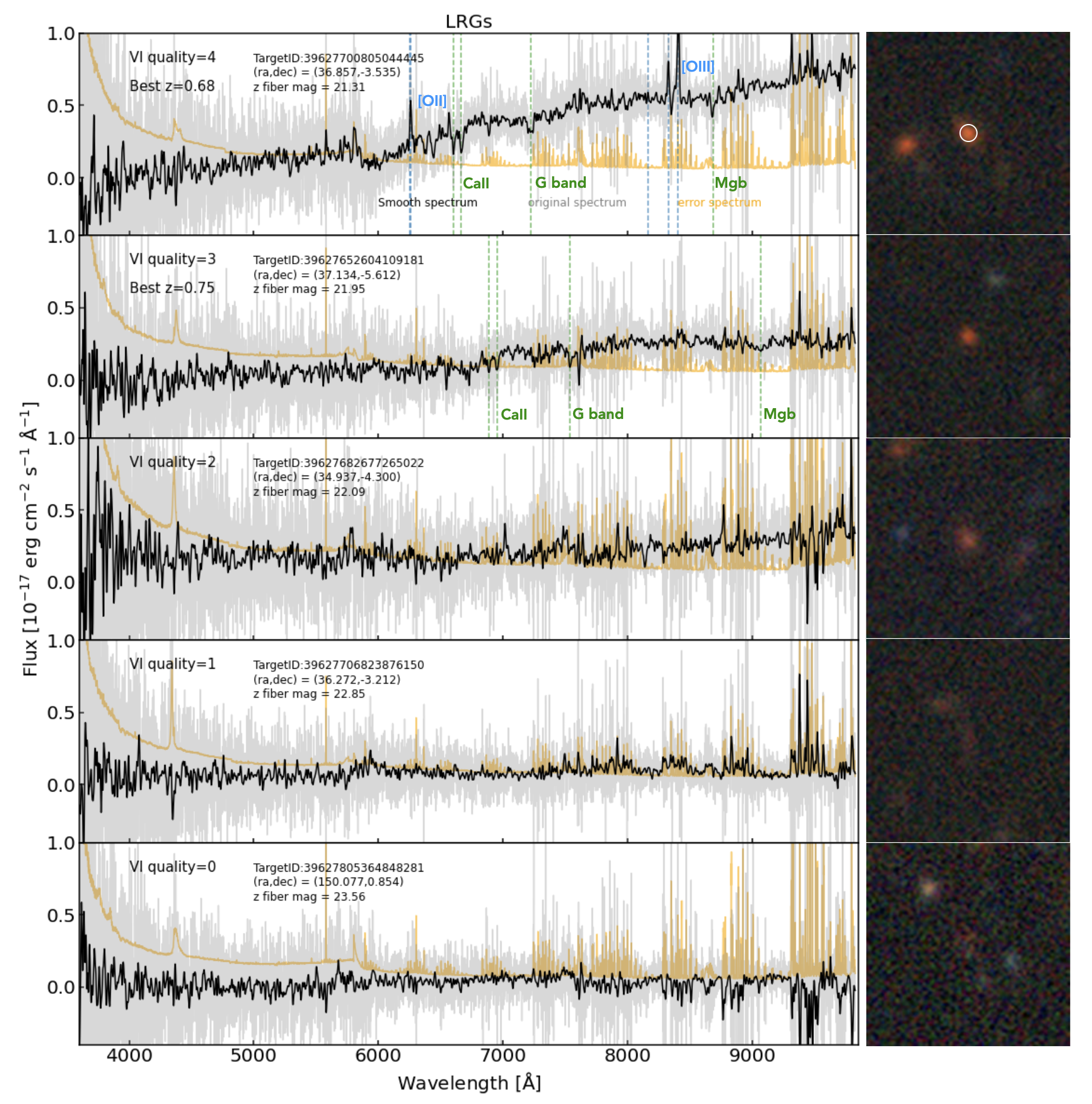}
\caption{{\tam Example of LRG spectra,} ordered by their VI quality values from top with VI quality 4 to bottom with VI quality 0. The quality 4 spectrum has multiple spectral lines, including CaII, G band, Mgb absorption lines and \oii emission lines. The quality 3 spectrum shows Ca II and G band absorption lines. Spectra with quality 2 and below do not have spectral features that can be robustly identifiable. Sky residuals can be observed at the wavelength regions $>8000\, \rm \AA$. The right panels show the DESI Legacy Survey images of the LRG targets. The spectra in grey, black and orange colors are the original observed galaxy spectrum, the smooth spectrum with a Gaussian filter, and the error spectrum, respectively.}
\label{fig:LRG_spectra}
\end{figure*}

\begin{figure*}
\center
\includegraphics[width=0.9\textwidth]{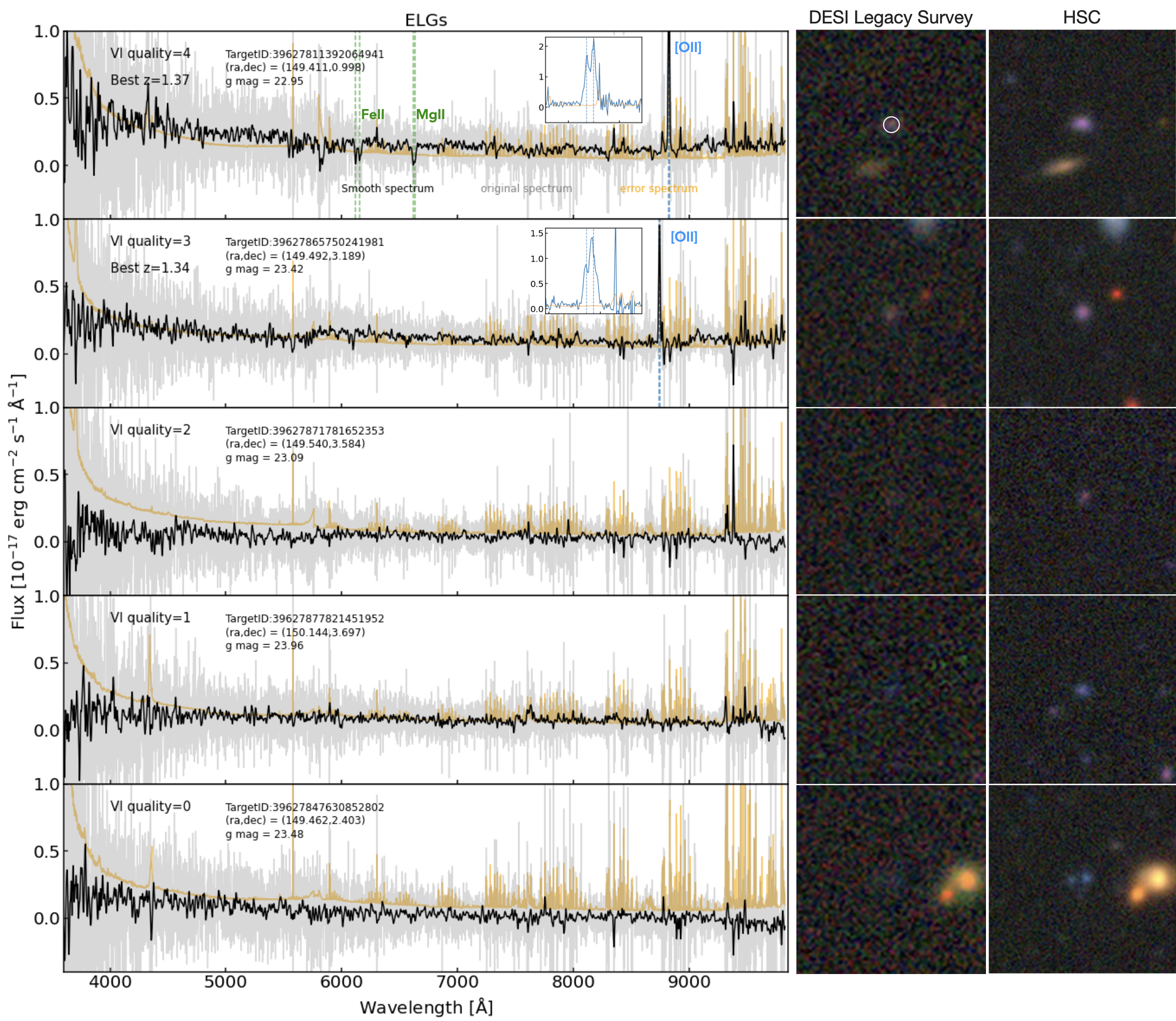}
\caption{{\tam Example of ELG spectra, ordered by their VI quality} values from top with VI quality 4 to bottom with VI quality 0. The quality 4 spectrum shows a resolved \oii doublet and interstellar absorption lines, including MgII and FeII. The quality 3 spectrum shows blended \oii emission lines. The insets show the detected \oii emission line profiles. The right two panels show the DESI Legacy Survey images and the Hyper Supreme-Cam images \citep{HSCDR2} of the ELG targets. The spectra in grey, black and orange colors are the original observed galaxy spectrum, the smooth spectrum with a Gaussian filter, and the error spectrum, respectively.} 
\label{fig:ELG_spectra}
\end{figure*}

\section{The procedure of visual inspection}
The main goals of visual inspection of DESI spectra are (1) to construct catalogs of different targets with redshift and source-type identifications verified by inspectors and (2) to assess the performance of the data pipeline and the redshift identifications from the \emph{Redrock} algorithm \citep{Redrock}. This VI step enables us to have additional redshift  classification information for each spectrum that does not entirely rely on the data pipeline and may be less sensitive to artifacts and systematic features in the data.

The general flow of the VI procedure is as follows:
\begin{enumerate}
    \item Each spectrum has at least two inspectors. 
    \item {\tam For each spectrum, inspectors report four key features:}
        \begin{enumerate}    
        \item Redshift;
        \item Quality of the redshift with 5 discrete values from 0 to 4 with 4 being the highest quality and 0 being the poorest quality;
        \item Type of the source, i.e.\ star, galaxy, or quasar;
        \item Any artifacts or systematic features in the spectrum;
        \item  Plus any extra comments, e.g., there are two objects in the spectrum.
        \end{enumerate}
       
    \item The outputs from the inspectors are compared. {\tam For each spectrum, the multiple VI reports are automatically merged into the final VI results, if the outputs of all the inspectors satisfy the following conditions:} 
    \begin{enumerate}    
        \item The difference of the redshifts, $dz$ from inspectors $i$ and $j$, is within $0.0033$ ($990 \rm \, km/s$), 
        \begin{equation}
            dz_{\rm VI}=\frac{|z_{{\rm VI,}i}-z_{{\rm VI,}j}|}{(1+z_{{\rm VI,}i})}<0.0033;
            \label{eq:dz_vi}
        \end{equation}
        \item the difference of the quality values is within~1, i.e. $|{\rm VI \ quality}_{i}-{\rm VI \ quality}_{j}|\leq 1$;
        \item The VI spectype from all of the inspectors is the same; 
        \end{enumerate}
    The final redshift is the mean of the redshifts, the final quality value is the mean of the quality values, and the final source type (best VI type) is the spectype identified by all of the inspectors.
    \item If any of the above conditions is not met, a merger inspects the spectrum again and determines the final VI results. Note that this merger might occasionally be one of the inspectors. 
\end{enumerate}
One of the key steps in the VI procedure is to assess the quality of the VI redshift. The criteria for the quality are: 
\begin{itemize}
    \item \textbf{Quality 4}: confident classification with two or more secure features, e.g., spectra with multiple emission lines and absorption lines;
    \item \textbf{Quality 3}: probable classification with at least one secure spectral feature and continuum or many weak spectral features, e.g., spectra with a strong emission line feature and weak Balmer series absorption lines; 
    \item \textbf{Quality 2}: possible classification with one strong spectral feature but unsure what it is;
    \item \textbf{Quality 1}: unlikely classification with some signal but features are unidentified;
    \item \textbf{Quality 0}: nothing there, no signal.
\end{itemize}
Figures~\ref{fig:BGS_spectra}, \ref{fig:LRG_spectra}, and \ref{fig:ELG_spectra} show example BGS, LRGs, and ELGs spectra respectively, of each of these different qualities. 

Given the above definitions, we consider the final VI redshift with an overall VI quality $\geq 2.5$ as a robust VI redshift. Sources with an overall VI quality $<2.5$ typically indicate that the corresponding spectra  are too shallow to detect spectral features or the sources are at redshift regions with spectral features beyond the wavelength coverage of the DESI spectrograph, e.g. galaxies at $z>1.62$ where the \oii doublet is beyond the wavelength coverage of the instrument. 

Finally, we summarize our dedicated tool, \emph{Prospect}, for performing the VI and our efforts of recruiting and training visual inspectors in the Appendix.

\section{Data and the catalogs}
The data used in this work was obtained by the DESI instrumentation with a 3.2 degree diameter field of view and 5020 robotic fiber positioners installed on the 4-meter Mayall Telescope at Kitt Peak National Observatory \citep[][]{instrument, Silber2022, Miller2022}. The DESI spectrograph covers the optical wavelength region from 3600~$\rm \AA$ to 9800~$\rm \AA$ with a spectral resolution ranging from $\sim 2000$ in the shortest wavelength channel to  $\sim 5000$ in the longest wavelength channel. In addition, the DESI targets are selected based on the photometric properties from the DESI Legacy Imaging Surveys \citep[][]{DR9,Zou2017,Dey2019}. The target selection procedure is summarized in \citet[][]{TS}. The pipelines used to automatically assign fibers to targets and used to plan and optimize the observational procedure are summarized in \citet[][]{Raichoor2022b} and \citet[][]{Schlafly2022}.

During the DESI SV observations, several fields were selected to have approximately $10$ times longer exposure time than the designed nominal exposure time of the DESI Main Survey \citep[][]{SVoverview}.  {\tam This is for detecting spectral features that might not be detected with the normal exposure time and for validating the exposure time of the DESI Main Survey. Because the observational conditions, such as airmass and galactic extinction of the field, vary, the exposure time for each observation was calibrated into the effective exposure time, $\rm T_{eff}$, that corresponds to the exposure time with airmass 1, zero galactic extinction, $1''.1$ seeing (full width half maximum), and zenith dark sky \citep[][]{SVoverview}.}   
In addition, to explore the target selection schemes that maximize the efficiency of the survey, galaxy targets were selected from a wider magnitude and color space during the SV observations than the range planned for the Main Survey. 

{\tam To support these analyses, we have conducted two visual inspection campaigns. During the first one, we visually inspected $\sim 17,000$ deep co-added spectra, which were based on an early version of the data reduction pipeline\footnote{Software release 20.12, including desispec 0.36.0 and redrock 0.14.4}, from SV observations, including BGS, LRG, and ELG targets. 
Through the visual inspection, we identified some spectral features due to instrument effects, calibration issues, or sky residuals which will be discussed in Section 5. These features prevented the inspectors from recovering the true redshifts of the sources.
Therefore, in the second campaign, we performed the visual inspection for sources with low VI quality values identified during the first one but with the spectra reduced by the latest version\footnote{Software release 22.2, including desispec 0.51.13 and redrock 0.15.4} of the data reduction pipeline. The sources included in the second campaign are 
\begin{enumerate}
    \item all SV BGS galaxies with VI quality$<=2.5$,
    \item all SV LRG galaxies with VI quality$<=2.5$, and 
    \item all ELGs in the Main Survey selection with VI quality$<=2.5$.
\end{enumerate}
Note that we only included ELGs that are in the Main Survey selection because the main goal of the analysis is to use the VI results to test the Main Survey target selection and performance. With this second campaign, we recovered a few percent of redshifts that were identified with low quality previously. In the final VI catalogs, we adopted the second campaign VI results for sources included in the second campaign and the first campaign VI results for the rest. The basic information of the VI fields and the statistics are summarized in Table~\ref{table:VI_info} and \ref{table:robust_redshift_fraction}.
} 
In the following, we describe the {\tam VI properties} of each target type. 

\begin{table*}
\centering
\caption{VI fields and catalogs of BGS, LRGs, and ELGs}
\label{table:VI_info} 
\centerline{
\tablewidth{0pt}
\hskip-2.5cm\begin{tabular}{c|c|c|c|c|c}
\hline
Targets & TILEID & Field name & Position & Effective exposure time$^{a}$ [s] & Number of sources \\
& &  &  (ra,dec) [deg] &  & (SV extended selection)  \\
\hline\hline

BGS & 80613 & Lynx & (106.740, 56.100) & $\sim$3000 &  2718  \\
\hline  
LRGs & 80605 & XMM-LSS & (36.448, -4.601) & $\sim$6900 & 1772  \\
     & 80609 & COSMOS & (150.120, 2.206) &$\sim$7800 & 1789 \\
\hline
ELGs & 80606 & XMM-LSS & (36.448, -4.501)  & $\sim$6700 & 3441 \\
     & 80608 & Lynx & (106.740, 56.200) & $\sim$15200 & 3430  \\
     & 80610 & COSMOS & (150.120, 2.306) & $\sim$9500 & 3444 \\
\hline
\end{tabular}
}
\tablecomments{ $^{a}$ These exposure times are $\sim7-10$ times longer than the nominal exposure times of the three tracers.}
\end{table*}

\begin{table*}[ht!]
\centering
\tablewidth{0pt}
\caption{Number of sources and robust VI redshift fraction} 
\hskip-2.5cm\begin{tabular}{c|c|c|c|c|c}
\hline
Targets & Selection & Number of sources & Number of robust VI redshift & Fraction & Density [$\rm deg^{-2}$] \\
\hline\hline
BGS & SV & 2718 & 2640 & 97.1\% & $\sim$2500 \\
    & Main bright & 1037 & 1033 & 99.6\% & $\sim$870 \\
    & Main faint & 509 & 509 & 100\% & $\sim$540 \\
\hline
LRG & SV & 3561 & 3513 & 98.7\% & $\sim$2120\\
& Main & 933 & 928 & 99.5\% & $\sim$610 \\
\hline
ELG & SV & 10315 & 7856$^{a}$ & 76.2\%  & $\sim$6500\\
& Main LOP & 2897 & 2474 & 85.4\% & $\sim$1950\\
& Main VLO & 668 & 664 & 99.4\%  & $\sim$450\\
\hline

\end{tabular}
\label{table:robust_redshift_fraction}
\tablecomments{ $^{a}$ We have conducted two VI campaigns as discussed in Section 3. In the second campaign, we only revisited the ELG sources that are in the main target selection. Therefore, the number of robust VI redshifts and fraction for ELG SV are lower limits.}
\end{table*}

\subsection{Bright galaxy survey (BGS)}
For the bright galaxy survey, we visually inspect spectra of 2718 BGS targets in one deep tile from the SV observations; {\tam the position of this field is listed in Table~\ref{table:VI_info}}. The cumulative effective exposure time of the spectra is about 3,000\,s, which is 10-15 times longer than the designed nominal exposure time, 150-200s, for bright galaxies. The top left panel of Figure~\ref{fig:VI_properties} shows the VI quality distribution of SV BGS targets in grey. Approximately 97\% of the BGS targets have VI quality greater than 2.5, i.e. their spectral features can be identified. The rest of the sources are mostly faint galaxies of approximately 23 magnitude in $r$-band and image artifacts due to bright stars (see the bottom panel of Figure~\ref{fig:BGS_spectra} as an example). 

In the Main Survey, two samples, BGS bright and BGS faint, are selected \citep{BGSTS}. 
The VI quality distributions for the BGS bright and faint targets satisfying the Main Survey selection are shown in black and purple respectively. {\tam The fraction of BGS bright and faint sources having VI quality $\geq 2.5$ is found to be $\sim$99.7\%.} We note that the BGS bright sample yields a lower fraction of VI robust redshifts (i.e. VI quality $\geq 2.5$) than the BGS faint sample. This is expected given that the BGS bright sample is selected based on a minimum set of simple selection criteria for cosmological analysis. On the other hand, the BGS faint sample is designed to use special color-magnitude selection criteria to maximize the redshift success rate of the sample \citep{BGSTS}.

The top right panel of Figure~\ref{fig:VI_properties} shows the redshift distributions of the BGS targets in the SV and Main Survey selections with robust VI redshifts. The median redshifts of the SV and Main Survey selections are 0.24 and 0.22, respectively. Both redshift distributions extend to $\sim0.7$. 

\subsection{Luminous red galaxies (LRGs)}
\begin{figure*}
\center
\includegraphics[width=1\textwidth]{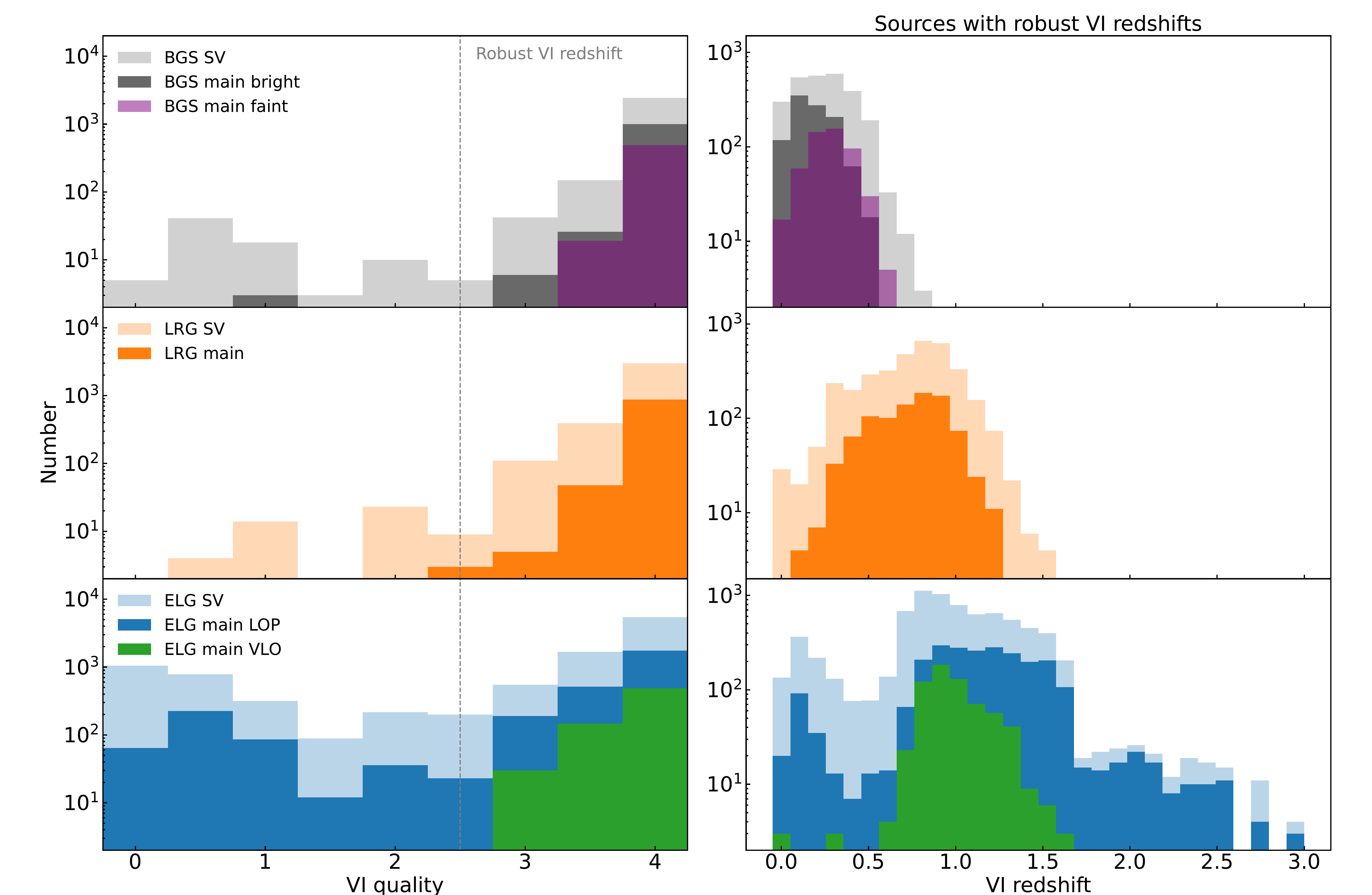}
\caption{\emph{Left:} Distributions of VI quality values of BGS, LRGs, and ELGs from top to bottom respectively. \emph{Right:} VI redshift distributions of BGS, LRGs, and ELGs with robust VI redshifts from top to bottom respectively. The faint color in each panel shows the distribution of sources selected in Survey Validation observations and the dark color shows the subset of those that are in the DESI Main Survey target selection.}
\label{fig:VI_properties}
\end{figure*}

For LRGs, we visually inspect 3561 spectra from two tiles. The cumulative effective exposure time is $\sim7000$s for both tiles, about 7 times longer than the designed nominal exposure time of $\sim 1000$s. The middle two panels of Figure~\ref{fig:VI_properties} show the VI qualities and redshifts of the LRGs in both SV (faint orange) and Main Survey (orange) selections. With the SV selection, $98.7\%$ of LRG spectra have VI quality $\geq2.5$. The remaining $1.3\%$ of the sources are mostly faint galaxies (22 magnitude in z-band) with no obvious spectral features detected. Similar to the BGS survey, a brightness cut has been adopted for the LRG selection in the Main Survey. With the Main Survey selection, $99.5\%$ of LRG spectra have VI quality$\geq2.5$. 
The median redshifts of the SV and Main Survey selections for LRGs in the VI catalog are 0.78 and 0.82, respectively. The redshift distributions both extend to $z \sim 1.2$.


\begin{figure*}
\center
\includegraphics[width=0.95\textwidth]{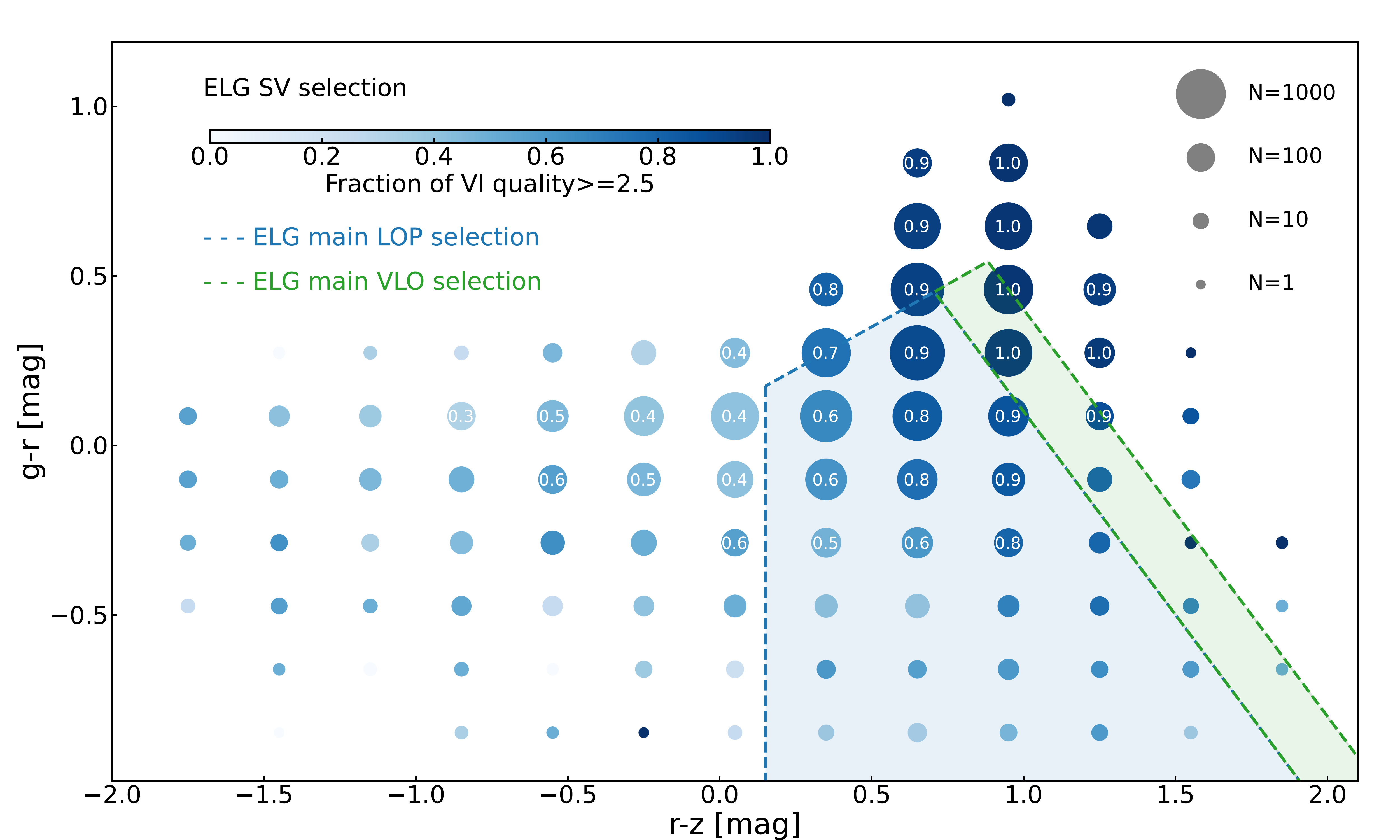}
\includegraphics[width=0.95\textwidth]{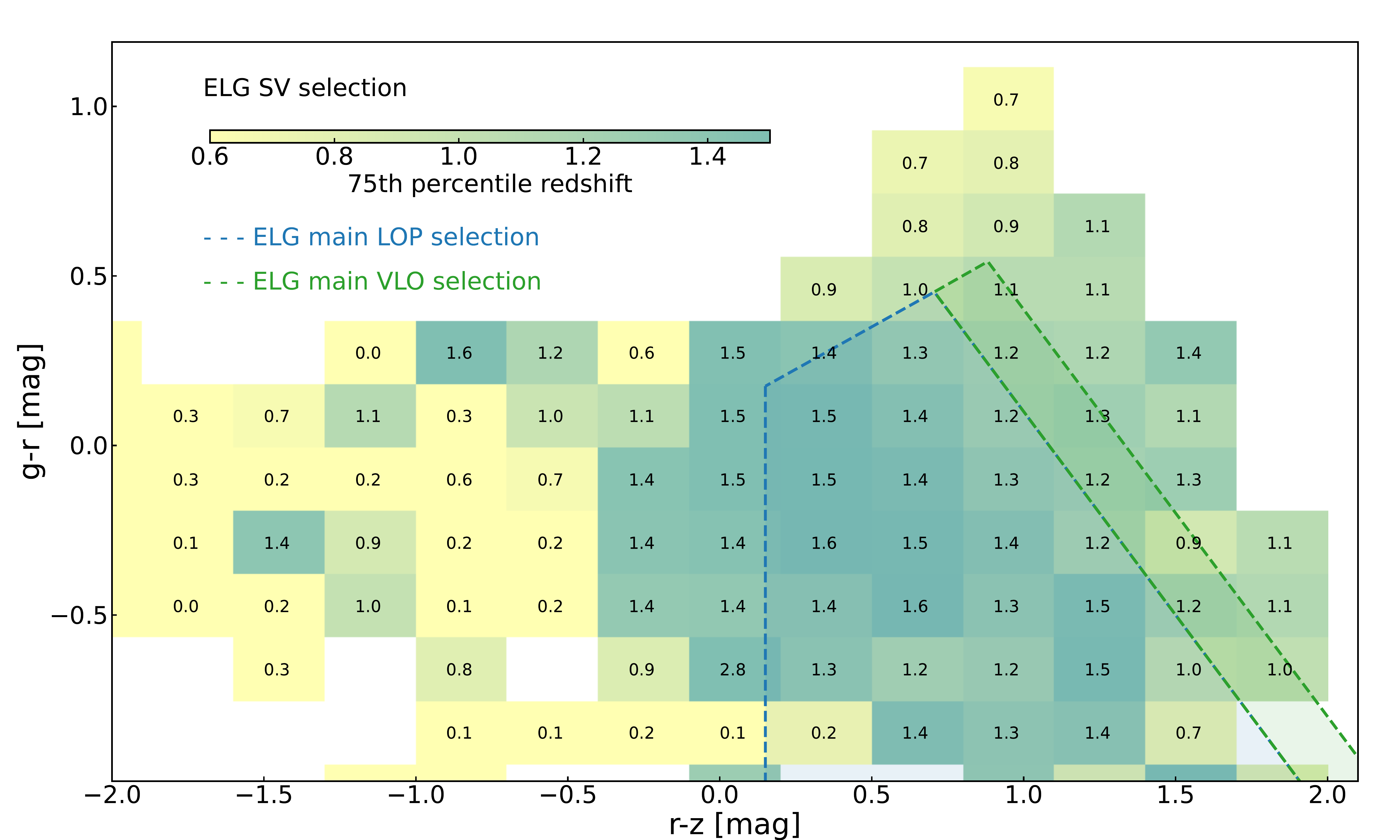}
\caption{ELG SV and Main Survey g-r and r-z selections. \emph{Top:} The fraction of ELGs with VI quality$\geq 2.5$. The color indicates the fraction and the size of the circles reflects the number of sources in the color-color bins. The dashed lines show the 
the LOP (blue) and VLO (green) 
color-color selection criteria 
used in the Main Survey. \emph{Bottom:} The 75th-percentile redshifts of ELGs with VI quality$\geq 2.5$ in each color-color bin. We note that for ELGs with $r-z<0$, most of their spectra ($>50\%$) do not have features that can be identified visually as shown in the top panel. The remaining ELGs with VI quality$\geq 2.5$ tend to be at low redshifts as shown in the bottom panel.}
\label{fig:ELG_selection}
\end{figure*}

\subsection{Emission line galaxies (ELGs)}
{\tam The DESI survey will target ELGs which have not been extensively studied up to this point, particularly at $z \gtrsim 1$.} Therefore, an extended selection of ELG targets is applied in the SV observations. Figure~\ref{fig:ELG_selection} shows the selection with $g-r$ and $r-z$ colors for the SV ELG targets. We visually inspect 10315 spectra of SV ELG targets spanning a wide range of this color-color space from three tiles with cumulative exposure times ranging from $\sim 7,000$ to $15,000$s. Among all the visually inspected ELG spectra, $\sim 75\%$ of them have VI quality $\geq2.5$. {\tam The remaining $25\%$ are mostly featureless spectra which are likely from galaxies at $z>1.62$ or galaxies with weak spectral features which are beyond the detection limit.} 
The VI quality and redshift distributions of ELGs are as shown in the lower panels of Figure~\ref{fig:VI_properties}.

The fraction of ELGs with VI quality $\geq2.5$ depends on the observed colors of the targets as shown in the top panel of Figure~\ref{fig:ELG_selection}. {\tam The colors in the figure indicate the fraction of sources with VI quality $\geq2.5$ in each bin.} As can be seen, galaxies with redder g-r and r-z colors {\tam in the top right corner have a larger fraction of visually-identified redshifts; these are lower-redshift galaxies with multiple emission lines in the spectral coverage of the DESI spectrograph.} 

The dashed lines in the figure are the boundaries in the color-color space for the Main Survey ELG selection. The ELG Main Survey consists of two selections, {\tam LOP and VLO} \citep{ELGTS}, which are the names for targeting bits \citep{TS}, indicating their fiber assignment priorities ('LOw Priority` and `Very-LOw priority') in the DESI survey.

The LOP selection, indicated by the blue shaded region in Figure~\ref{fig:ELG_selection}, is designed to select ELGs at $1.1<z<1.6$ (the bottom right panel of Figure~\ref{fig:VI_properties}) with a
target density of approximately 1950~$\rm deg^{-2}$. 
The motivation of this selection is to maximize the number of ELG spectra with recoverable redshifts at $1.1<z<1.6$, a redshift range that is mostly probed by ELGs via \oii emission lines, with the {\tam designed Main Survey exposure time} \citep{ELGTS}. This LOP ELG is the main ELG sample designed for cosmological measurements. 
On the other hand, the VLO selection (the green shaded region in Figure~\ref{fig:ELG_selection}) includes galaxy targets with redder colors than the LOP targets. Therefore, the VLO selection includes ELGs at slightly lower redshifts $0.6<z<1.6$, with a higher redshift recovery rate than the LOP selection. Moreover, the VLO ELG sample has a lower observation priority than the LOP ELG sample in the main DESI survey. {\tam We note that while galaxies with $g-r>0.6$ have high density and high VI recoverable redshift rate, they are at redshifts lower than the main redshift range of interest probed by the DESI ELG samples.} 


 The bottom panel of Figure~\ref{fig:ELG_selection} shows the redshifts of galaxies with VI quality $\geq2.5$ at the 75th percentile in each color bin. As can be seen, the two ELG Main Survey selections encompass the parameter space which yields recoverable redshifts within the desired range and at relatively high target density. This is also reflected in the bottom right panel of Figure~\ref{fig:VI_properties}, which shows that this main selection includes ELGs mostly at $0.6<z<1.6$ with about $10\%$ of sources being low-z galaxies $z<0.6$ and quasars ($z>1.6$) . 
In the LOP and VLO Main Survey selections, $\sim 85\%$ and $\sim99\%$ of ELG targets have VI quality $\geq2.5$ respectively (see Table~\ref{table:robust_redshift_fraction} for details).

\begin{figure*}
\center
\includegraphics[width=0.6\textwidth]{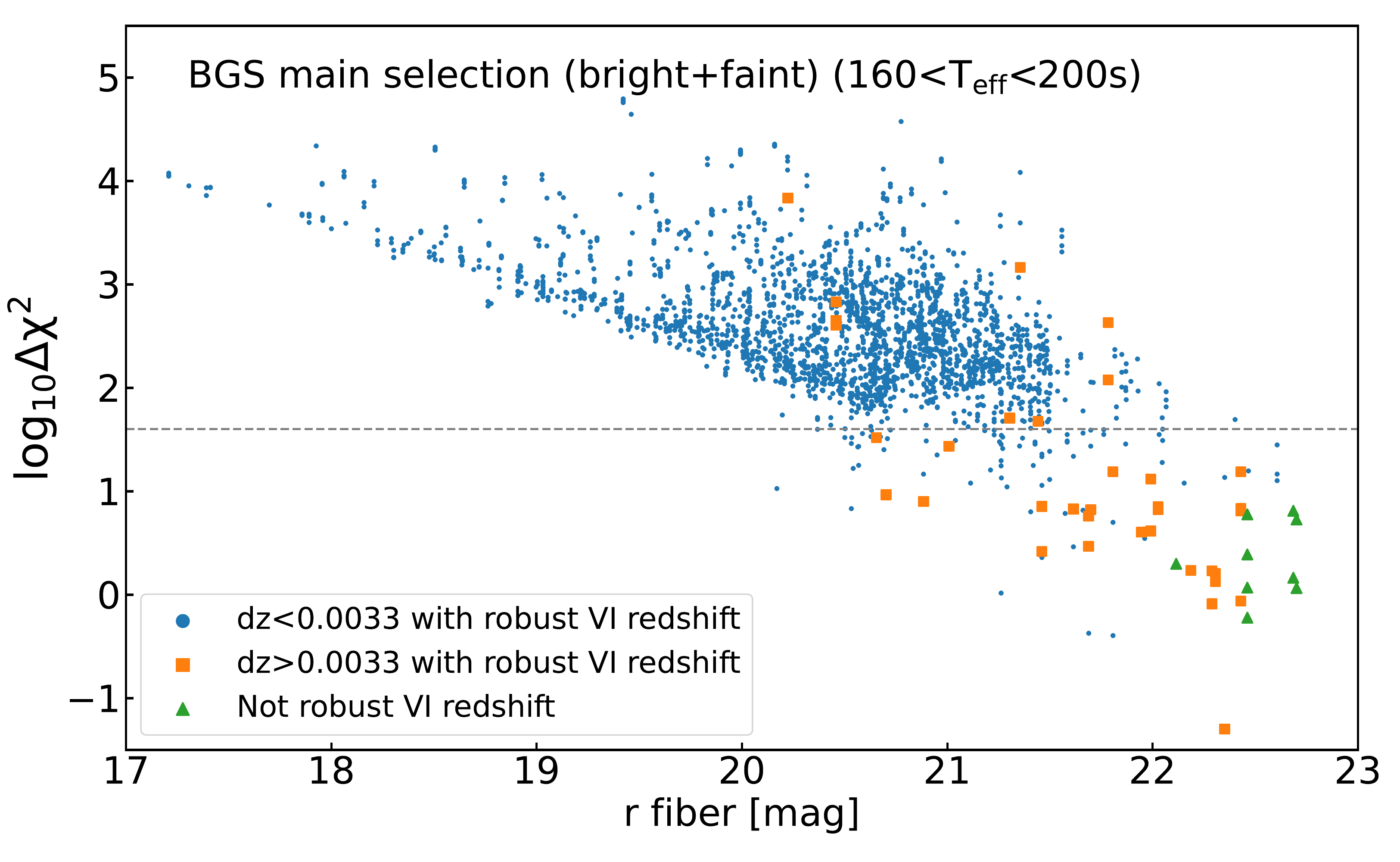}
\includegraphics[width=0.6\textwidth]{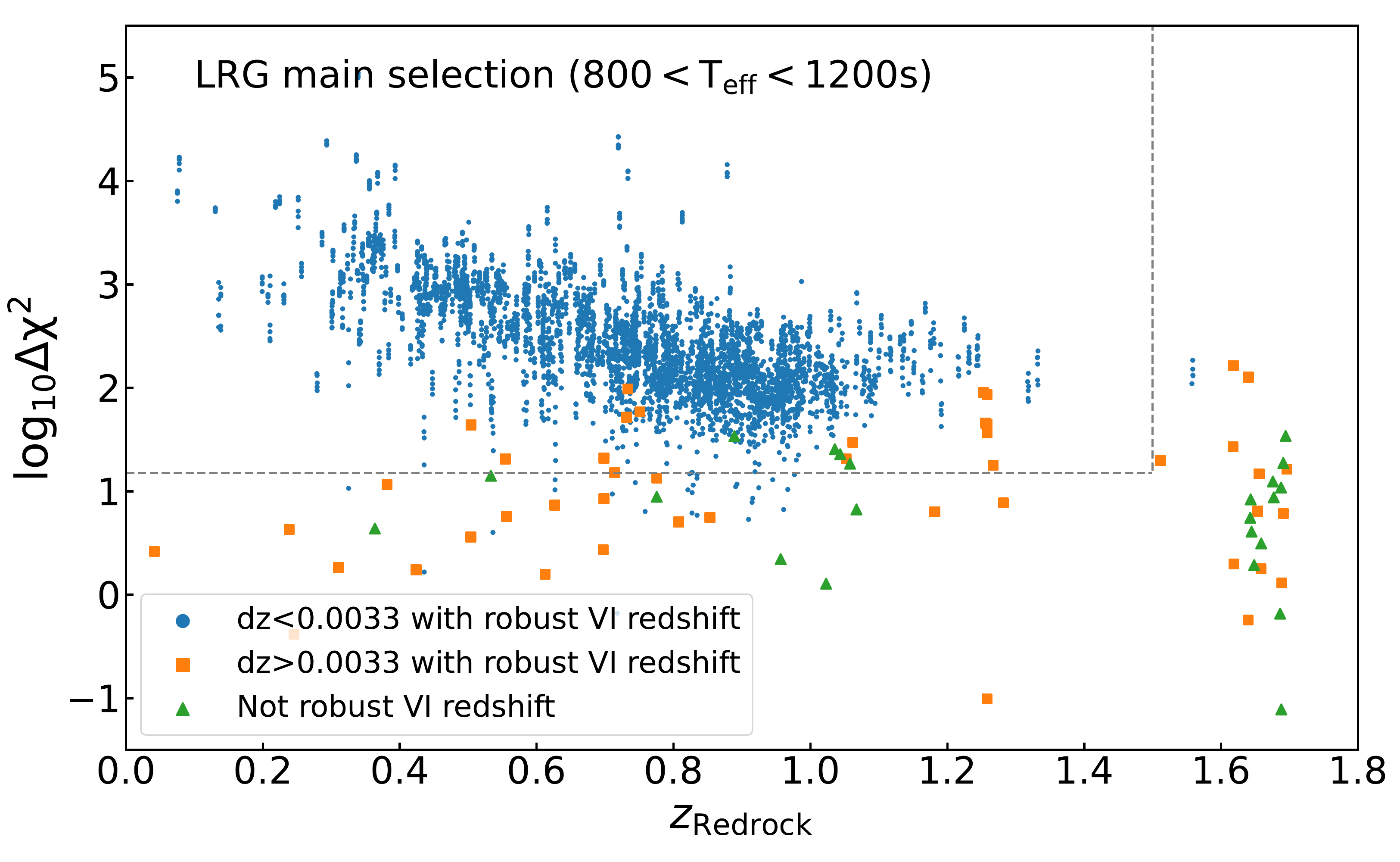}
\includegraphics[width=0.6\textwidth]{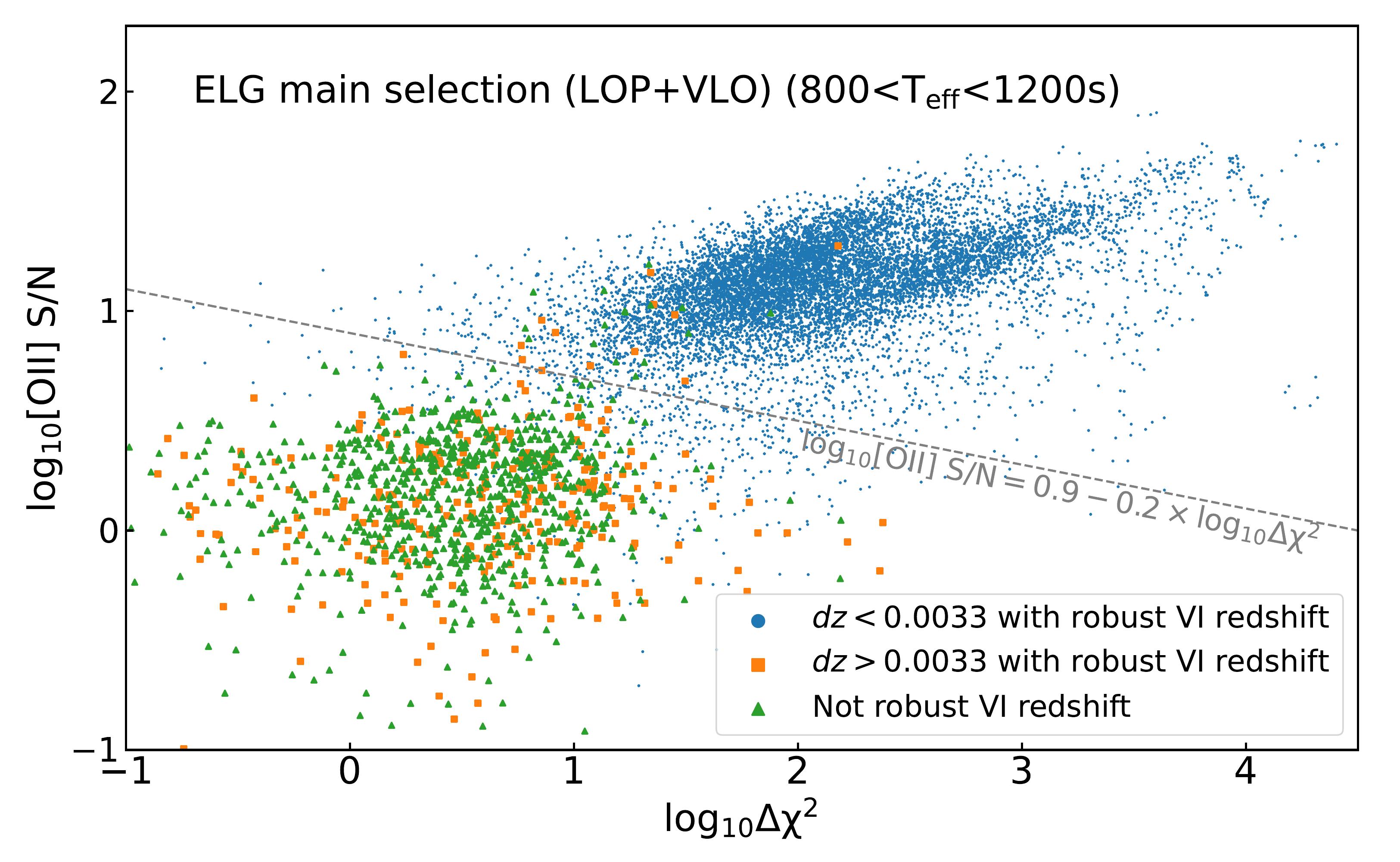}
\caption{$\Delta \chi^{2}$ distributions as a function of r-band fiber magnitudes of BGS (top) and \emph{Redrock} redshifts of LRGs (middle). $\Delta \chi^{2}$ and \oii emission line S/N plane of ELGs (bottom). The blue data points show the distributions of sources with robust VI redshifts (VI quality $\geq$2.5) and with \emph{Redrock} redshifts being consistent with the VI redshifts ($dz<0.0033$). The orange squares show the distributions of  sources with robust VI redshifts but with \emph{Redrock} redshifts being inconsistent with the VI redshifts ($dz>0.0033$). The green triangles show the distributions of sources with no robust VI redshifts. The grey dashed lines indicate the adopted selection criteria 
to include the majority of objects with redshifts successfully identified by \textit{Redrock} for BGS, LRGs, and ELGs. The criteria are discussed in Section 4.2.}
\label{fig:LRG_selection_z_vs_deltachi2}
\end{figure*}

\section{Validating the DESI survey design with VI catalogs}
We use the VI galaxy catalogs obtained from DESI spectra with long exposure times to validate the performance of the algorithm, \emph{Redrock} \citep{Redrock}, that DESI adopts for recovering the redshifts of galaxies at the designed nominal exposure times, i.e. effective exposure time $\sim 1000$s for ELGs \citep{ELGTS} and LRGs \citep{LRGTS} and $\sim 180$s for bright galaxies \citep{BGSTS}. For those galaxies with deep co-added spectra, we use all the data from individual exposures to produce several spectra with similar exposure times to the designed exposure time of the Main Survey. For ELGs and LRGs, we use spectra with $\rm 800<T_{eff}<1200s$ and for BGS, we use spectra with $\rm 160<T_{eff}<200s$.
In this way, one galaxy target has multiple redshift measurements from the \emph{Redrock} algorithm, and we can then compare the redshift measurements from the pipeline with the VI redshift to assess the performance of the DESI observation and data pipeline. 

There are two key quantities from \emph{Redrock} used in this data assessment:
\begin{itemize}
    \item \textbf{Redrock redshift}: the best-fit redshift of the spectrum from \emph{Redrock},
    \item \textbf{$\Delta \chi^{2}$}: the difference between the $\chi^{2}$ value of the 2nd best-fit model and the $\chi^{2}$ value of the 1st best-fit model, $\chi^{2}_{2nd} - \chi^{2}_{1st}$. A larger $\Delta \chi^{2}$ means that the 1st best-fit model better describes the data than the 2nd best-fit model. The $\Delta \chi^{2}$ parameter is an important parameter for identifying robust redshifts as described below. 
\end{itemize}

\subsection{Redshift recovery rate}
We first estimate the redshift recovery rate, i.e.\ the fraction of robust VI redshifts (VI quality $\geq 2.5$) that are recovered by \emph{Redrock} from spectra with Main Survey exposure times: 
\begin{equation}
    {\rm Redshift \ recovery \ rate} = \frac{N(dz<0.0033 \ \rm \& \ VI \ quality \geq 2.5)}{N(\rm VI \ quality \geq 2.5)}
\end{equation}
where the redshift difference ($dz$) between the VI redshifts and Redrock redshifts is defined as      
\begin{equation}
    dz=\frac{|z_{\rm VI}-z_{\rm Redrock}|}{(1+z_{\rm VI})}.
    \label{eq:dz_assessement}
\end{equation}
In this calculation, we exclude sources identified as stars by the \emph{Redrock} pipeline. 

The redshift recovery rates for the Main Survey selections are
\begin{itemize}
    \item BGS bright: 98.1$\pm$0.3\%, 
    \item BGS faint: 99.4$\pm$0.3\%,
    \item LRG main: 98.9$\pm$0.2\%,
    \item ELG LOP: 93.8$\pm$0.2\%,
    \item ELG VLO: 97.5$\pm$0.3\%.
\end{itemize}
These high redshift recovery rates indicate that the Main Survey exposure times are sufficient for obtaining spectra with spectral features that can be automatically detected and identified by \emph{Redrock}. 

We note that there are two main reasons for sources with $dz>0.0033$: 
(1) some of the spectra are not deep enough to robustly detect spectral features and therefore \emph{Redrock} can not identify the true redshifts, and (2) some of the spectra are affected by artifacts introduced during the calibration process. Those spectral artifacts drive the best-fit models and the corresponding redshifts. In Section 5, we will show {\tam some examples of spectra with artifacts} that were found during this analysis and VI process.

\begin{figure}[h]
\center
\includegraphics[width=0.495\textwidth]{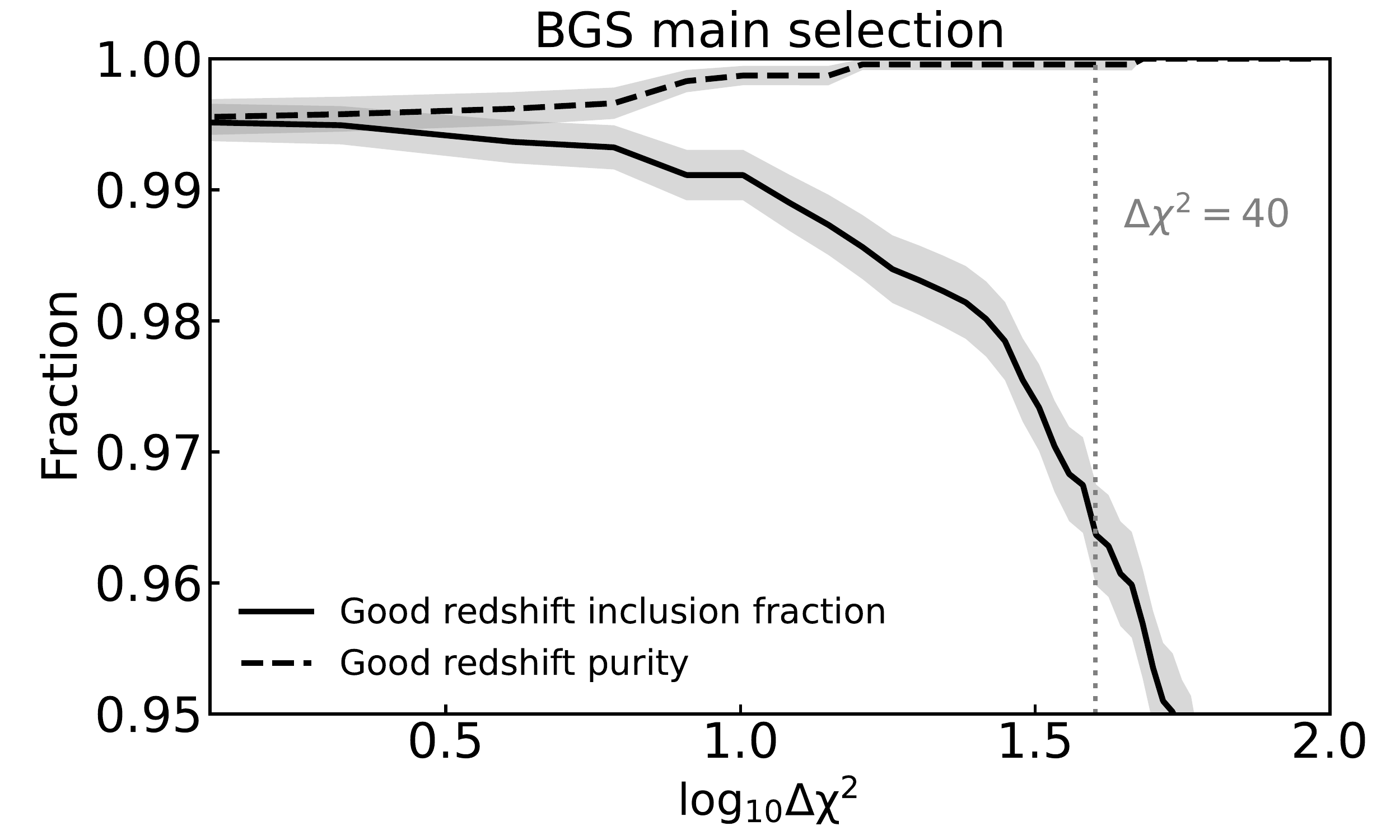}
\includegraphics[width=0.495\textwidth]{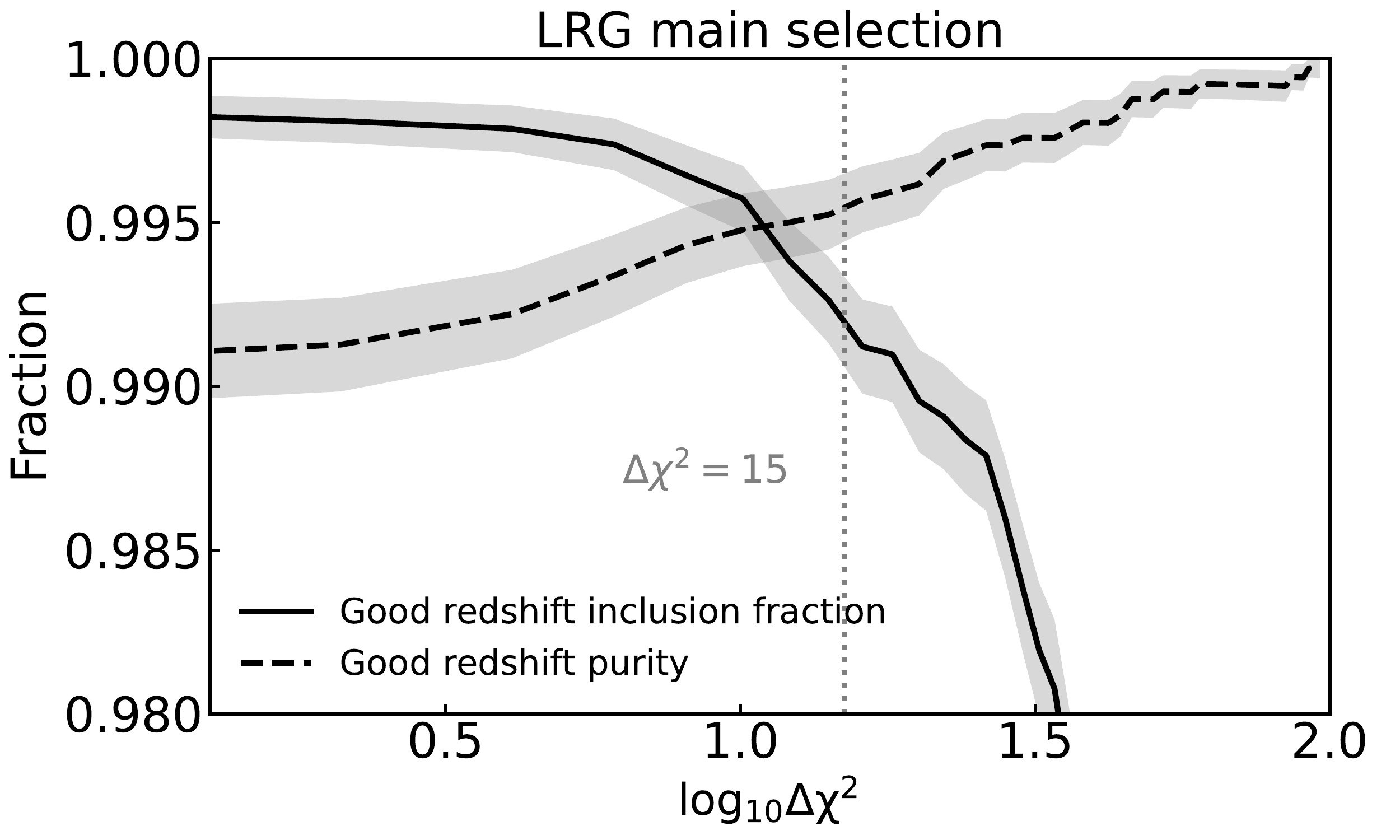}
\includegraphics[width=0.495\textwidth]{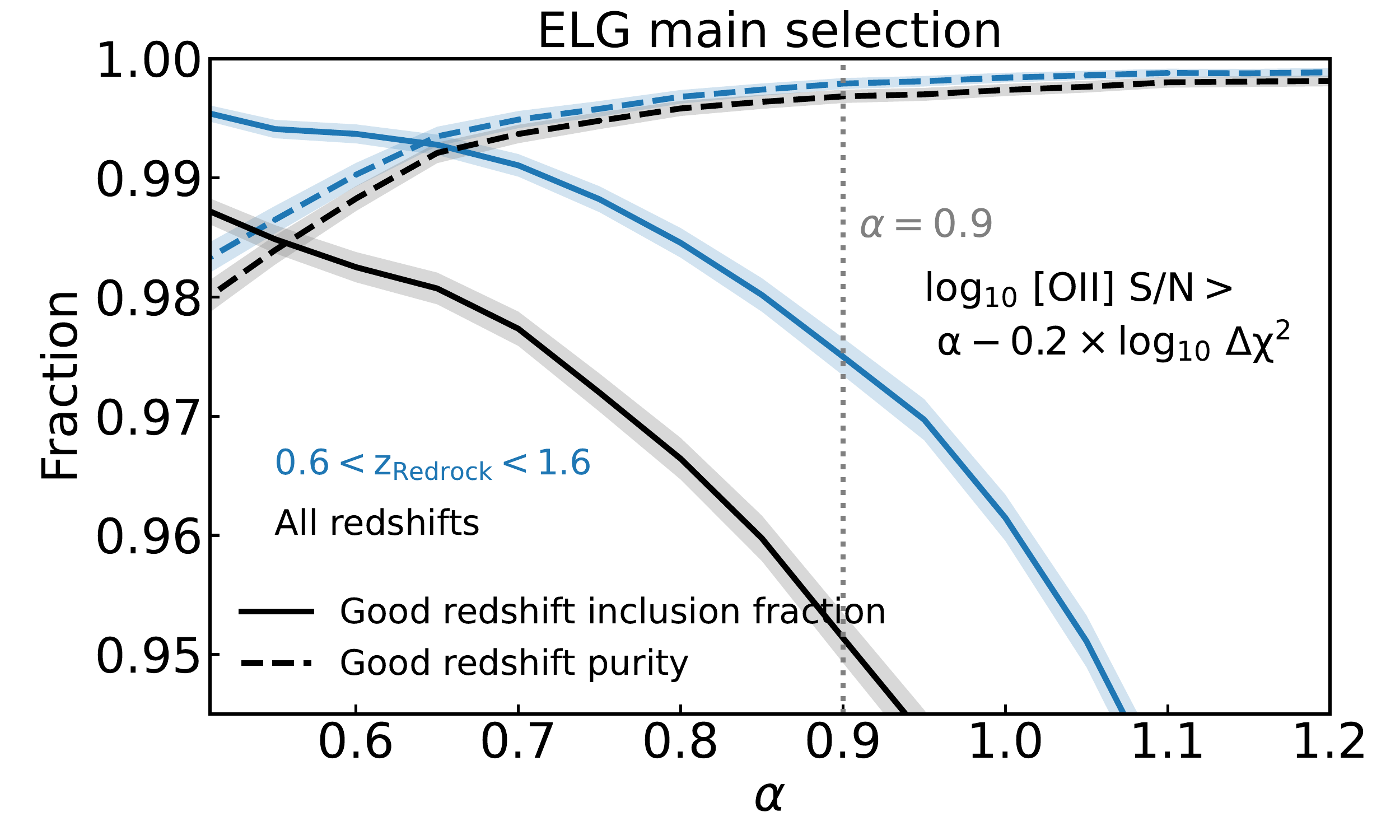}
\caption{Good redshift inclusion fraction and good redshift purity measurements as a function of $\Delta \chi^{2}$ for BGS main survey selection, including BGS bright and BGS faint samples (top) and LRG main selection (middle) and as a function of $\alpha$ for ELG main survey selection, including LOP and VLO samples (bottom). The solid lines show the measurements of the good redshift inclusion fraction and the dashed lines show the measurements of the good redshift purity. For ELGs, the black curves show the values of ELGs in the Main Survey selection at all redshifts and 
the blue curves show the results of the Main Survey ELG sample with \emph{Redrock} redshifts between 0.6 and 1.6. The shaded region indicates the $1\sigma$ uncertainty for  binomial distributions.} 
\label{fig:BGS_LRG_inclusion}
\end{figure}

\subsection{$\Delta \chi^{2}$ selections}
{\tam Most spectra obtained from the DESI Main Survey will not have VI validated redshifts. Therefore, one of the key tasks during the SV phase is to use the VI catalogs to identify the combination of \emph{Redrock} $\Delta \chi^{2}$ values and other parameters from DESI spectra that we can use as a selection criterion in order to (1) include most of objects with redshifts successfully identified by \emph{Redrock} and (2) exclude most objects with incorrect redshifts. To this end, we define the following quantities:}
\begin{align}
    & {\rm Good \ redshift \ inclusion \ fraction}\equiv \\ \nonumber
    & \frac{N(dz<0.0033 \ \& \ {\rm VI \ quality}\geq 2.5 \ \& \ {\rm criteria _{passed} })}{N(dz<0.0033\  \& \ {\rm VI \ quality} \geq 2.5)}
\end{align}
and 
\begin{align}
    & {\rm Good \ redshift \ purity}\equiv \\ \nonumber
    & \frac{N(dz<0.0033 \ \& \ {\rm VI \ quality} \geq 2.5 \ \& \ {\rm criteria _{passed} })}{N({\rm criteria _{passed} })}.
\end{align}
The good redshift inclusion fraction indicates the fraction of \emph{Redrock} recovered redshifts that would be included in the sample if we apply a selection criterion, e.g. $\Delta \chi^{2}>\rm threshold$. A higher value means higher efficiency of obtaining 
sources with correct redshifts from the automatic \emph{Redrock} redshifting algorithm.

\begin{table*}
\centering
\caption{Metrics for assessing the DESI performance} 
\begin{tabular}[t]{c|c|c|c|c|c|c}
\hline
Targets & Selection &  Good redshift  & Good redshift   & Redshift    & Redshift  & Redshift  \\
 &  & inclusion fraction  &  purity  & efficiency  & precision & accuracy\\

\hline\hline
 
BGS &  bright & 97.2$\pm$0.4\% & 100\%  & 87.0$\pm$0.8\% ($0.01<z<0.5$) & 9.2$\pm$0.5 km/s & 6.5$\pm$1.7 km/s  \\
    &  faint  & 94.8$\pm$0.8\% & 99.9$\pm$0.1\% & 87.7$\pm$1.1\% ($0.01<z<0.5$) & 8.5$\pm$0.7 km/s &  \\
\hline
LRG & main & 99.2$\pm$0.1\% & 99.6$\pm$0.1\% & 86.4$\pm$0.5\%  ($0.4<z<1.1$) & 36.7$\pm$0.6 km/s &  -3.0$\pm$3.3 km/s \\
\hline
ELG & LOP & 94.3$\pm$0.3\% & 99.6$\pm$0.1\%  & 67.0$\pm$0.5\% ($0.6<z<1.6$) &  6.8$\pm$0.1 km/s & -1.0$\pm$0.4 km/s \\
      &     &  &  & 35.5$\pm$0.5\% ($1.1<z<1.6$)& \\
ELG & VLO & 97.9$\pm$0.3\% & 99.9$\pm$0.1\% &  93.4$\pm$0.5\% ($0.6<z<1.6$) & 8.4$\pm$0.2 km/s &\\
      &     &  &  & 20.4$\pm$0.8\% ($1.1<z<1.6$) & &\\      
\hline      
\end{tabular}
\label{table:metrics}
\end{table*}
On the other hand, the good redshift purity indicates the fraction of objects with validated \emph{Redrock} redshifts that pass the selection criterion, e.g., $\Delta \chi^{2}>\rm threshold$ , compared to all \emph{Redrock} redshifts passing the selection criterion. A higher value means that the selected sample has a higher fraction of sources with correct redshifts. 
{\tam We emphasize that the $\Delta \chi^{2}$ value is the key quantity that helps us assess the reliability of redshifts for most of the spectra obtained by the DESI survey. Therefore, it is crucial to estimate the good redshift purity to ensure the success of redshift determinations.}

To assist the interpretation of these two quantities, we show the $\Delta \chi^{2}$ values as a function of observed magnitudes of BGS in the top panel of Figure~\ref{fig:LRG_selection_z_vs_deltachi2}. The blue data points show objects with $dz<0.0033$ and VI quality$>=2.5$ (good \emph{Redrock} redshifts), and the orange squares and green diamonds indicate objects with $dz>0.0033$ and VI quality$>=2.5$, and VI quality$<2.5$ (not robust VI redshift) respectively. 
As can be seen, most systems with confirmed \emph{Redrock} redshifts tend to have high $\Delta \chi^{2}$ values and most systems with incorrect redshifts tend to have low $\Delta \chi^{2}$ values. In the following, we describe the current adopted selection for each target. The corresponding inclusion fraction and purity values are summarized in Table~\ref{table:metrics}. Sources identified as stars by the \emph{Redrock} pipeline are excluded in the calculation.

\textbf{Results for BGS ---}
{\tam To estimate the redshift purity and inclusion fraction, we include redshift measurements from $\it{Redrock}$ in the selection criteria \citep[see also][]{BGSTS}: $z_{\rm Redrock}<0.6$}.
The top panel of Figure~\ref{fig:BGS_LRG_inclusion} shows the good redshift inclusion fraction and good redshift purity for BGS as a function of $\Delta \chi^2$. As expected, the good redshift inclusion fraction decreases with increasing $\Delta \chi^2$. Similarly, the redshift purity increases with $\Delta \chi^{2}$ since most of the inconsistent redshifts are excluded in the samples with a high $\Delta \chi^{2}$ threshold. 
We note that the redshift purity of the BGS sample is always $>99\%$ and the good redshift inclusion fraction is also very high ($>96\%$). 

As can be seen, adopting different $\Delta \chi^{2}$ thresholds for BGS will yield samples with different purity and inclusion fraction values as well as the number of sources in the sample. 
Currently, the DESI survey adopts the $\Delta \chi^{2}=40$ threshold for BGS as indicated by the vertical grey dashed line in Figure~\ref{fig:BGS_LRG_inclusion}.
This threshold will yield high redshift purity samples for both BGS bright and faint galaxies (nearly $100\%$). 

\textbf{Results for LRGs ---}
Similarly to BGS, for LRGs, we include the redshift measurements from $\it{Redrock}$ in the selection criteria \citep[see also][]{LRGTS}: $z_{\rm Redrock}<1.5$.
The middle panel of Figure~\ref{fig:BGS_LRG_inclusion} shows the good redshift inclusion fraction and good redshift purity of LRGs as a function of $\rm log_{10} \Delta \chi^{2}$ and the vertical dashed line shows the current adopted value $\rm \Delta \chi^{2}=15$.
This adopted selection is shown by the dashed line in the middle panel of Figure 6, yielding a sample of LRGs with the inclusion fraction and purity values $>99\%$.

\begin{figure*}
\center
\includegraphics[width=0.95\textwidth]{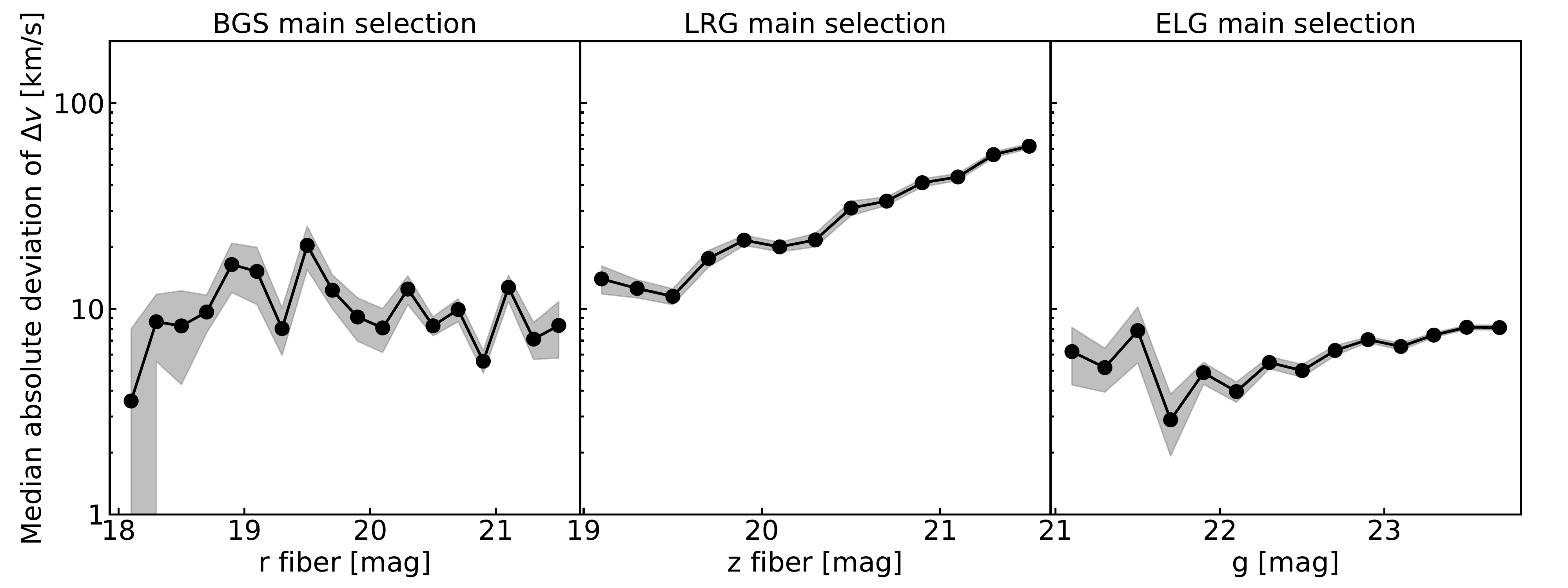}
\caption{Median absolute deviation of the distributions of redshift difference $\Delta v$ from pairs of \emph{Redrock} redshift measurements for BGS (left), LRGs (middle), and ELGs (right). Here we use the MAD values scaled to the standard deviation and divided by $\sqrt{2}$ to obtain the value representing the random uncertainty of one redshift measurement. The shaded region indicates the $1\sigma$ uncertainty estimated by bootstrapping the sample 500 times.} 
\label{fig:dv}
\end{figure*}

\begin{figure}
\center
\includegraphics[width=0.45\textwidth]{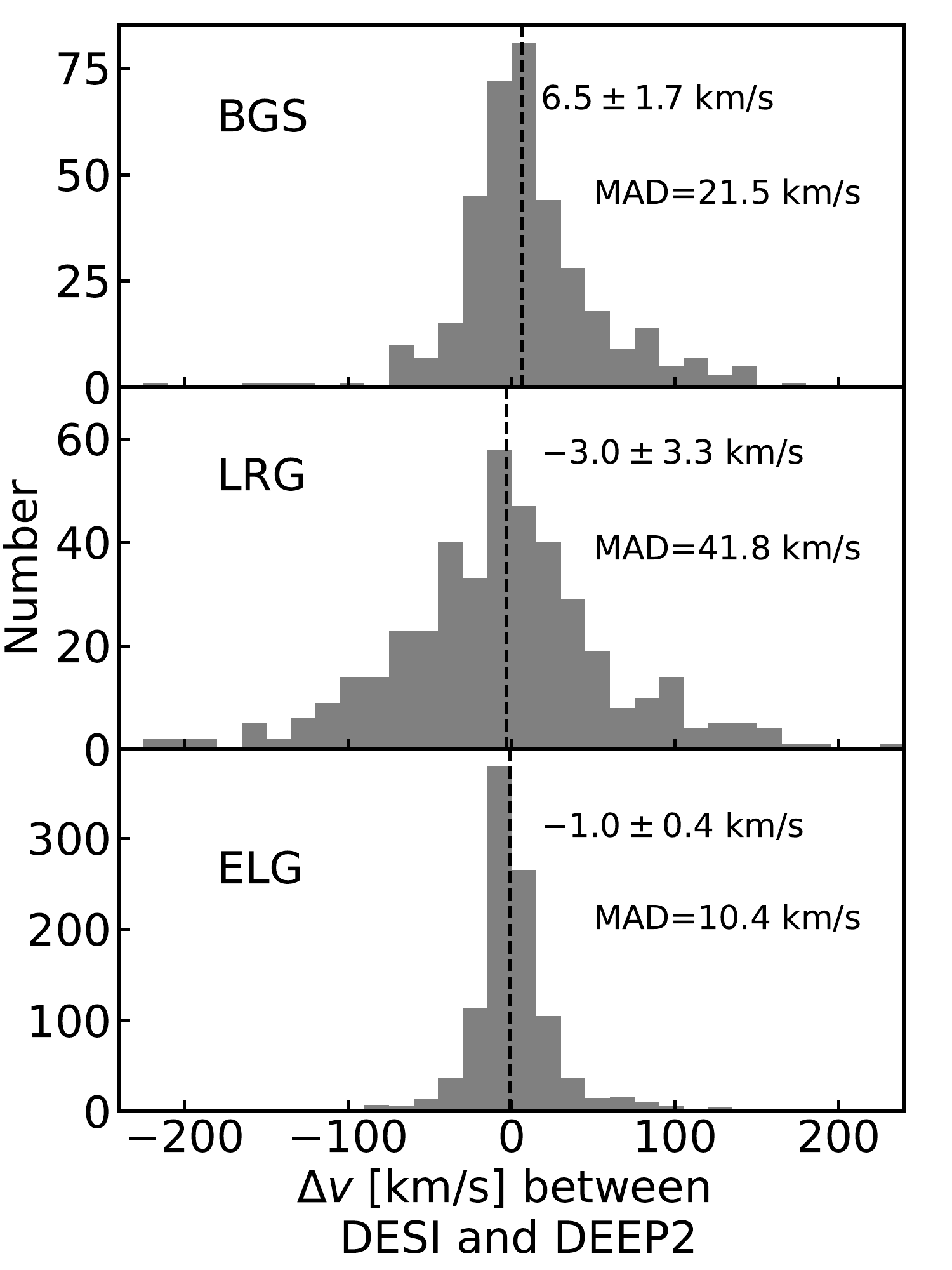}
\caption{Distributions of redshift differences between DESI \emph{Redrock} redshifts and DEEP2 redshifts for BGS (top), LRGs (middle), and ELGs (bottom). Dashed lines and velocity values show the median redshift differences between the samples.}
\label{fig:dv_systematics}
\end{figure}

\textbf{Results for ELGs ---} For ELGs, we find that a joint constraint of the signal to noise ratio of \oii emission line doublet ([OII] S/N) and  $\Delta \chi^{2}$ is more efficient for selecting high inclusion fraction and redshift purity sample than a single constraint from $\Delta \chi^{2}$ \citep{ELGTS}. 
This is because a fraction of the
ELG spectra only have \oii emission lines without any detectable continuum flux. These spectra yield low $\Delta \chi^{2}$ but with \oii emission lines that can be robustly detected. The emission line properties are obtained via a line-fitting algorithm used in the DESI pipeline \citep[Section 6.3][]{ELGTS}. The bottom panel of Figure~\ref{fig:LRG_selection_z_vs_deltachi2} shows that the successful \emph{Redrock} recovered redshifts (blue data points) and the inconsistent redshifts of ELGs (green diamonds and orange squares) can be separated by a function, 
\begin{equation}
    \rm log_{10} \ [OII] \ S/N = \alpha - \beta \times log_{10}\ \Delta \chi^{2}.
\end{equation}
{\tam To illustrate how the good redshift inclusion fraction and the redshift purity values depend on the parameters, in the bottom panel pf Figure~\ref{fig:BGS_LRG_inclusion}, the black lines show the values of ELGs at all redshifts as a function of $\alpha$ with a fixed $\beta=0.2$ and the blue lines show the results of the main survey ELG sample with \emph{Redrock} redshifts between 0.6 and 1.6. We note that the difference between the good redshift inclusion fraction values of ELGs at all redshifts and ELGs with $\rm 0.6<z_{Redrock}<1.6$ is due to 
the fact that most ELGs with $\rm 0.6<z_{Redrock}<1.6$ have strong \oii emission lines which allow them to pass the selection criterion.} 
Currently, we adopt the following selection for ELGs \citep{ELGTS}:
\begin{itemize}
    \item \textbf{ELGs}: $\rm log_{10} \ [OII] \ S/N > 0.9 - 0.2 \times log_{10}\ \Delta \chi^{2}$.
\end{itemize}
{\tam For ELGs at all redshifts, this selection yields an ELG LOP sample with redshift purity $\sim99.6\%$ and good redshift inclusion fraction $\sim 94.3\%$ and an ELG VLO sample with redshift purity $\sim99.9\%$ and good redshift inclusion fraction $\sim 97.9\%$. For ELGs with $\rm 0.6<z_{Redrock}<1.6$, this selection yields an ELG LOP sample with redshift purity $\sim99.8\%$ and good redshift inclusion fraction $\sim 97.3\%$ and an ELG VLO sample with redshift purity $\sim99.9\%$ and good redshift inclusion fraction $\sim 98.2\%$.}

\subsection{Redshift efficiency}
Previously, we have used the redshift inclusion fraction and the redshift purity to quantify the properties of the DESI data with given selection criteria, demonstrating that we can obtain samples of BGS, LRGs, and ELGs with high purity values ($>99\%$). 

We now define a quantity which allows us to address the following question: for all the galaxy targets that were observed by DESI with nominal exposure times, what is the fraction of sources that 
{\tam
\begin{enumerate}
    \item pass the selection criteria,
    \item have their true redshifts recovered by Redrock ($dz < 0.0033$ and VI quality $>= 2.5$), and 
    \item have redshifts within the desired ranges for cosmological measurements $z_{\rm min}<z_{\rm VI}<z_{\rm max}$.
\end{enumerate}
}
We define the redshift efficiency, 
\begin{equation}
    {\rm Redshift \ efficiency}= \frac{N_{\rm passed}}{N_{\rm all}},
\end{equation}
where $N_{\rm passed}$ is the number of spectra that satisfy the three conditions listed above, and $N_{\rm all}$ is the total number of observed spectra of a given target type. 

The values of the redshift efficiency for BGS, LRGs, and ELGs are summarized in Table~\ref{table:metrics}. 
We find that for BGS with $z<0.5$ and LRGs with $0.4<z<1.1$, the redshift efficiency is higher than $86\%$. For ELG LOP and VLO samples with $0.6<z<1.6$, the redshift efficiency values are $\sim 67\%$ and $\sim 93\%$, respectively. Together with the target density and the fiber assignment rate, these redshift efficiency values are used to estimate the final galaxy densities with redshifts. {\tam The final estimations of galaxy densities are described in \citet{SVoverview} and the target selection papers of each tracer, \citet[][]{BGSTS}, \citet[][]{ELGTS}, and \citet[][]{LRGTS}.}

\subsection{Redshift precision}
We now estimate the random error (1$\sigma$) of the \emph{Redrock} redshift measurements of galaxies selected in the Main Survey.
To do so, we make use of the redshift measurements of the multiple individual spectra of the same target from the SV observation listed in Table 1.
Each spectrum has an effective exposure time similar to the effective exposure time of the Main Survey and the corresponding redshift passes the selection criteria described in Sec 4.2. 
We compare the redshift differences between all the pairs of \emph{Redrock} redshift measurements, e.g. 
$dv_{ij} = c (z_{i}-z_{j})/(1+z_{i})$
 and then use the median absolute deviation (MAD), an estimator which is less sensitive to outliers, to estimate the dispersion of the $dv$ distribution.  
Note that the MAD value is scaled to the standard deviation to reflect the 1$\sigma$ error. Finally, we divide the MAD value by $\sqrt{2}$ because the dispersion of $dv$ includes the errors of two redshift measurements:
\begin{equation}
\rm Reported \ MAD \ value = MAD \times 1.4828 / \sqrt{2}    
\end{equation}
Figure~\ref{fig:dv} shows the final MAD values of BGS, LRGs and ELGs as a function of galaxy brightness. As shown in the figure, the random errors of redshift measurements of BGS, ELGs, and LRGs are about $10$, $10$, and $40$ km/s respectively. 

\begin{figure*}
\center
\includegraphics[width=0.95\textwidth]{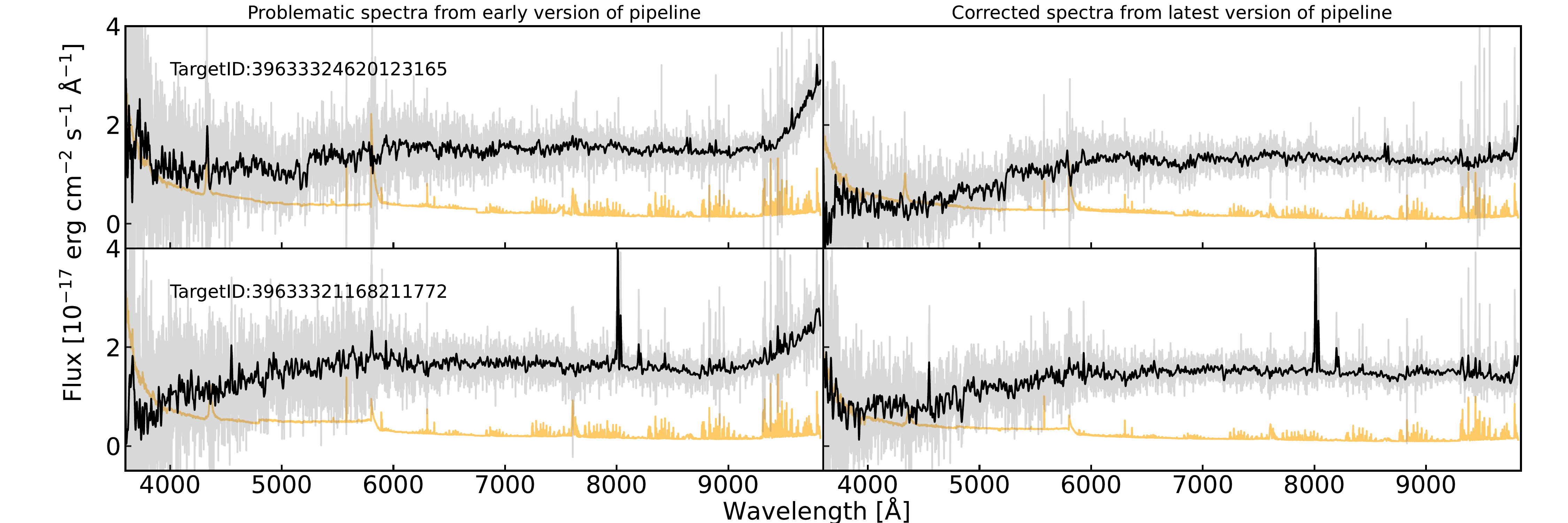}
\caption{Example of problematic spectra (left) identified by VI and the corresponding corrected spectra (right). The upturn of flux in the spectra shown in the left panel is due to flux contamination from the next fiber targeting a bright star.}
\label{fig:fiber_cross_talk}
\end{figure*}
\begin{figure}
\center
\includegraphics[width=0.45\textwidth]{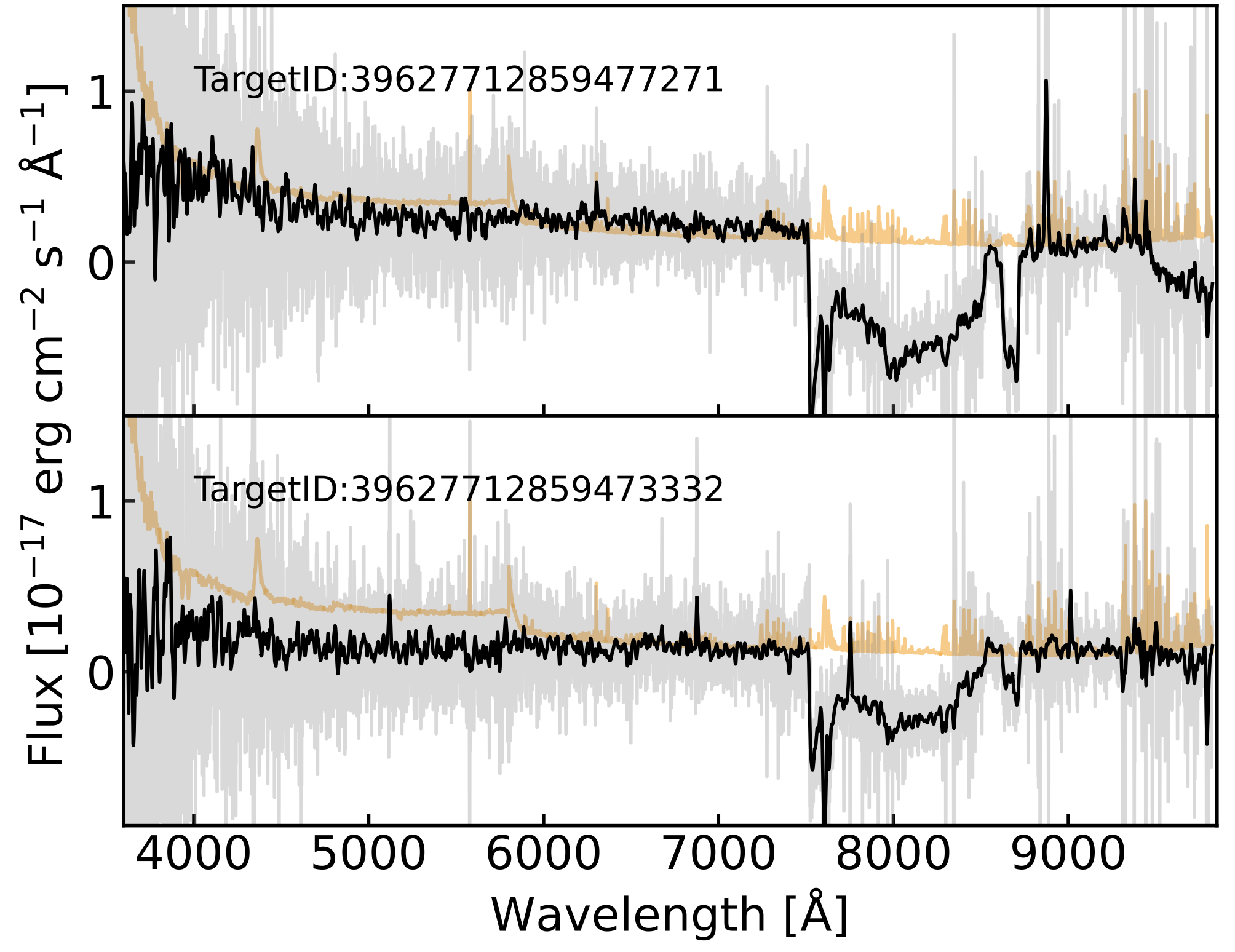}
\caption{Example of problematic spectra due to a CCD issue. }
\label{fig:fiber_2670}
\end{figure}

\subsection{Redshift accuracy}
In addition to the random error of redshift measurements, we investigate the  \emph{Redrock} redshift systematic inaccuracy of galaxies. While it is impossible to obtain the true redshifts of DESI galaxies and to assess the intrinsic redshift systematic uncertainty, we can compare the redshift measurements of DESI with the measurements from other surveys with similar spectral resolution and independent pipelines and obtain \emph{relative} redshift systematic uncertainty between two instruments. 

To do so, we use the DEEP2 redshift measurements \citep{Newman2013} with redshift quality $\geq 3$ and cross-match DEEP2 galaxies with galaxies observed by DESI. Here we use the galaxy data from DESI One-Percent Survey \citep{SVoverview} to increase the number of redshift pairs. We only use the redshift measurements that pass the selection criteria for BGS, LRGs and ELGs. 
In total, there are approximately 1800 DEEP2-DESI redshift pairs, including about 400 BGS, 400 LRGs and 1000 ELGs. 
Figure~\ref{fig:dv_systematics} shows the distributions of redshift differences between DEEP2 and DESI for BGS, LRGs, and ELGs. The vertical dashed lines show the median values of the redshift differences: 6.5 km/s, -3.0 km/s, and -1.0 km/s for BGS, LRGs, and ELGs respectively.
The median absolute deviation of the redshift differences
includes the redshift uncertainties of DESI and DEEP2 measurements. The MAD values are larger than the random error of the DESI redshift measurements as shown in Figure~\ref{fig:dv} due to the contribution from the DEEP2 redshift uncertainties \citep{Newman2013}.

This result indicates that the relative redshift systematic uncertainty between DESI and DEEP2 is within 10 km/s. Given that DEEP2 and DESI use two independent instruments and pipelines, the redshift consistency between the two suggests that the measured redshifts from the two surveys are close to the true redshifts with systematic redshift uncertainty $<10$ km/s.

\section{Assisting the development of the DESI pipeline}
Visually inspecting the DESI spectra is a crucial step in identifying unforeseen problems in the dataset and to further help the development and improvement of the data pipeline. To demonstrate the importance of VI, here we list two major issues which were identified by VI:
\begin{itemize}
    \item \textbf{Artificial excess of flux in z-band:} While visually inspecting spectra processed by an early version of the pipeline, we identified that 1-2\% of BGS spectra showed a non-physical increase in flux in the z-band, as shown in the left panel of Figure~\ref{fig:fiber_cross_talk}. This non-physical feature 
    led to erroneous estimates of the redshift using the Redrock algorithm as it simulates a strong $\rm 4000 \AA$ break.
    After investigation, we identified that this feature is due to the contamination of flux from adjacent fibers targeting bright stars. The data pipeline was updated to properly model this effect \citep{pipeline}. The right panel of Figure~\ref{fig:fiber_cross_talk} shows the spectra of the same targets as shown in the left panel but processed by the latest version of the pipeline. This pipeline improvement reduces the catastrophic failure rate of BGS redshifts by a factor of three. 
    
    \item \textbf{Spectra affected by a CCD issue:} During the VI process, we identified spectra with unexpected spectral features around $8000 \rm \, \AA$ as shown in Figure~\ref{fig:fiber_2670}, causing Redrock to fit incorrect redshifts.
    About 0.5\% of spectra reduced by an early version of the DESI pipeline have such features. 
    We find that these problematic spectra tend to be obtained from a certain set of fibers and are due to a region of CCD with issues, which were not identified and properly masked. The spectra affected by this problem are now masked by the data pipeline and will not be used in the redshift catalogs \citep{pipeline}. 
\end{itemize}
These results demonstrate that the visual inspection process 
is crucial for identifying unexpected problematic features which can affect the quality of the data product. Without VI in the early development of the project, 
it would be difficult to find such problems in the dataset.

\section{Rare objects identified by VI}
\begin{figure*}
\center
\includegraphics[width=0.95\textwidth]{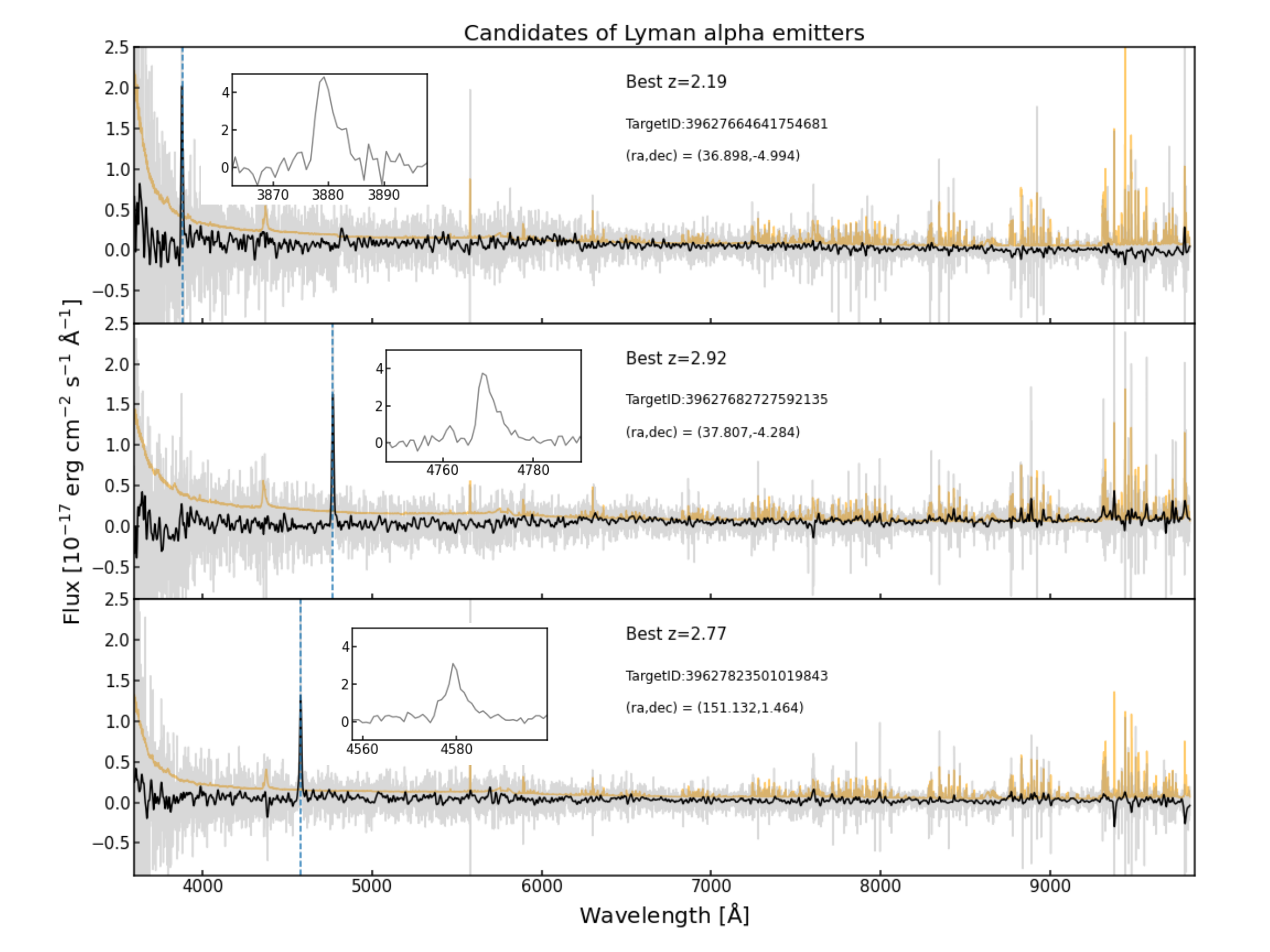}
\caption{Example spectra of Lyman $\alpha$ emitter candidates. The blue vertical dashed lines indicate the potential Lyman $\alpha$ emission lines. The spectra in grey, black, and orange colors are the original observed spectrum, the smooth spectrum with a Gaussian filter, and the uncertainty spectrum, respectively.
The inset of each panel shows the asymmetric Lyman $\alpha$ emission line.}
\label{fig:lya_spectra}
\end{figure*}

\begin{figure*}
\center
\includegraphics[width=0.95\textwidth]{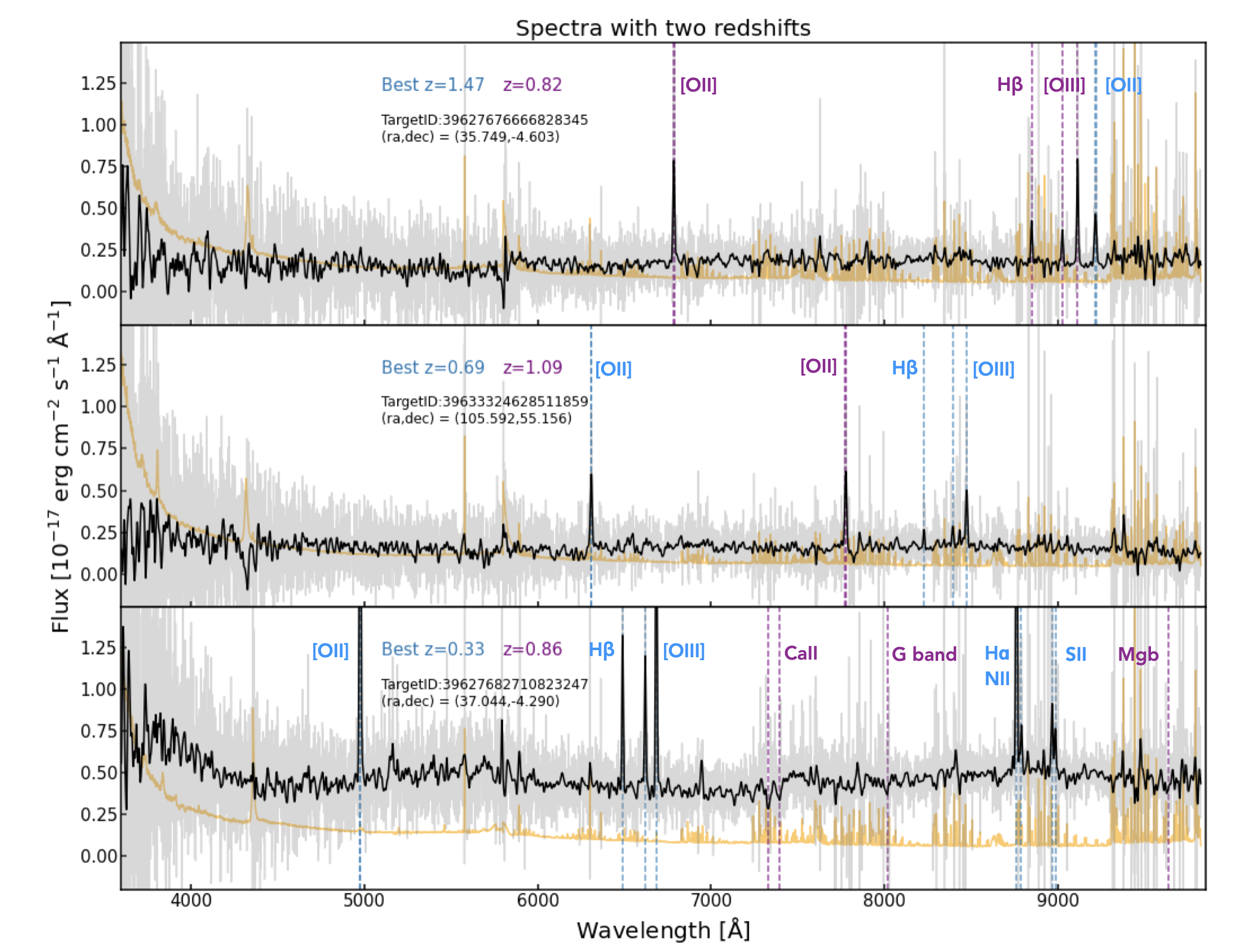}
\caption{Example spectra with two redshifts. Spectral features of the two redshifts are indicated by the purple and blue vertical dashed lines. The spectra in grey, black, and orange colors are the original observed spectrum, the smooth spectrum with a Gaussian filter, and the uncertainty spectrum, respectively. }
\label{fig:twoz_spectra}
\end{figure*}

In addition to revealing nonphysical spectra, the visual inspection process also helped us identify unexpected sources that live in the same parameter spaces as the DESI targets. Here we show two types of rare sources that were identified during the VI process:
\begin{itemize}
    \item \textbf{Candidates of Lyman $\alpha$ emitters:} We find about 60 spectra that show one strong, asymmetric emission line feature at about 4000-5000 $\rm \AA$ in observed wavelengths. These spectra are consistent with the spectra of Lyman $\alpha$ emitters, a star-forming galaxy population that produces a strong Lyman $\alpha$ emission line \citep[e.g.,][]{Leclercq2017, Ouchi2020}. Figure~\ref{fig:lya_spectra} shows examples of these spectra. This detection demonstrates that it is possible to use DESI to detect Lyman $\alpha$ emitters and potentially use them as mass tracers at $z>2$ for probing structure formation \citep[e.g.,][]{Hill2008}.
    \item \textbf{Spectra with two redshifts:} We also identify about 200 spectra comprised of two objects at two distinct redshifts, examples of which are shown in Figure~\ref{fig:twoz_spectra}. The purple dashed lines correspond to the spectral features of one galaxy and the blue dashed lines correspond to the features of another galaxy. These sources are lensing system candidates \citep[e.g.,][]{Bolton2008,Brownstein2012,Talbot2018}. If confirmed, these sources can be used to probe the stellar mass and dark matter mass around foreground galaxies, especially for the halos of star-forming galaxies at $z>1$ which are rarely explored via strong lensing.  
\end{itemize}
We note that for most of the Lyman $\alpha$ emission spectra, \emph{Redrock} fails to obtain the correct redshifts due to the fact that this unusual spectral feature is not included in the templates. We plan to add these VI identified spectra of potential Lyman $\alpha$ emitters into the \emph{Redrock} templates. By doing so, future versions of \emph{Redrock} may automatically identify similar spectra. {\tam For the spectra with two redshifts identified by VI, {\it Redrock} reports one of the redshifts that has dominated spectral features.}

\section{Conclusions}
During the Survey Validation of the DESI survey, we visually inspected approximately 17,000 DESI spectra of galaxy targets and 
thereby compiled VI redshift catalogs. 
These catalogs enable us to perform various statistical measurements for characterizing the performance of the DESI observational system. Here we summarize our main results:
\begin{itemize}

    \item We show that the redshifts of $97\%$ of BGS, $99\%$ of LRGs and $76\%$ of ELGs with long-exposure spectra observed in the SV observations can be identified from the VI process. These VI redshift catalogs are used to help define the selections for the Main Survey. In the Main Survey selections, the redshifts of $>99\%$ of BGS and LRGs can be visually identified and the redshifts of $\sim 85\%$ of LOP ELGs and $\sim 99\%$ of VLO ELGs can be visually identified. 
    
    \item We use the VI redshift catalogs to characterize the performance of the DESI Main Survey design and the \emph{Redrock} algorithm. The results show that with the current selected criteria combining $\Delta \chi^{2}$ and other parameter values, the DESI survey can obtain samples of BGS, LRGs, and ELGs with purity $>99\%$.
    
    \item Utilizing the spectra of the same targets obtained from multiple independent exposures, we quantify the precision of the redshift measurements from the \emph{Redrock} algorithm. The results show that the precision of the redshift measurements of BGS and ELGs is about 10 km/s and that of LRGs is about 40 km/s.
    
    \item We empirically test the accuracy of the redshift measurements from the \emph{Redrock} algorithm by comparing the \emph{Redrock} redshifts to the redshift measurements of the DEEP2 survey, using galaxies observed in both surveys. The results show that the systematic redshift differences from the two surveys are 
    less than 10 km/s for BGS, LRGs, and ELGs. 
    
    \item Via the VI process, we identify spectra with non-physical features. These identifications help us reveal and fix issues in both the software and hardware of the early DESI data pipeline that were not anticipated. In addition, we find spectra of rare astronomical objects, including candidates of Lyman $\alpha$ emitters and spectra with two redshifts as potentially strong lensing candidates. These spectra illustrate the scientific potential offered by the DESI dataset. 
\end{itemize}
These results illustrate the utility and necessity of performing systematic visual inspections of astronomical datasets, especially during the early phase of sky surveys. As demonstrated in this work, the VI redshift information is essential for testing the redshift measurements from the pipeline and for identifying both spurious features in the reduced spectra and unexpected astronomical phenomena. 

These VI catalogs will have a lasting impact. They will be used to perform the same statistical measurements every time the DESI pipeline is upgraded throughout the whole DESI operation. By doing so, we can confirm the stability of the DESI pipeline and assess the improvement quantitatively. {\tam The catalogs can also be used to train novel machine learning models of detecting and identifying spectral features.}
Finally, with this experience of visually inspecting DESI spectra, we recommend that upcoming spectroscopic surveys, such as PFS \citep{PFS}, SDSS-V \citep{Kollmeier}, 
4MOST \citep{4most} and MOONS \citep{Moons}, deploy similar VI efforts during the survey operations. 

\section*{Data Availability}
All data points shown in the figures are available at Zenodo \url{https://doi.org/10.5281/zenodo.6618765}.

\vspace{\baselineskip}
TWL is supported by the Ministry of Science and Technology (MOST 111-2112-M-002-015-MY3), the Ministry of Education, Taiwan (MOE Yushan Young Scholar grant NTU-110VV007 and NTU-110VV007-2), National Taiwan University research grants (NTU-CC-111L894806, NTU-111L7318, and NTU-112L7302), and NSF grant AST-1911140. 
DMA acknowledges the Science Technology and Facilities Council (STFC) for support through grant code ST/T000244/1.

This research is supported by the Director, Office of Science, Office of High Energy Physics of the U.S. Department of Energy under Contract No. DE–AC02–05CH11231, and by the National Energy Research Scientific Computing Center, a DOE Office of Science User Facility under the same contract; additional support for DESI is provided by the U.S. National Science Foundation, Division of Astronomical Sciences under Contract No. AST-0950945 to the NSF’s National Optical-Infrared Astronomy Research Laboratory; the Science and Technologies Facilities Council of the United Kingdom; the Gordon and Betty Moore Foundation; the Heising-Simons Foundation; the French Alternative Energies and Atomic Energy Commission (CEA); the National Council of Science and Technology of Mexico (CONACYT); the Ministry of Science and Innovation of Spain (MICINN), and by the DESI Member Institutions: \url{https://www.desi.lbl.gov/collaborating-institutions}.

The authors are honored to be permitted to conduct scientific research on Iolkam Du’ag (Kitt Peak), a mountain with particular significance to the Tohono O’odham Nation.

\appendix
\begin{figure*}
\center
\includegraphics[width=1\textwidth]{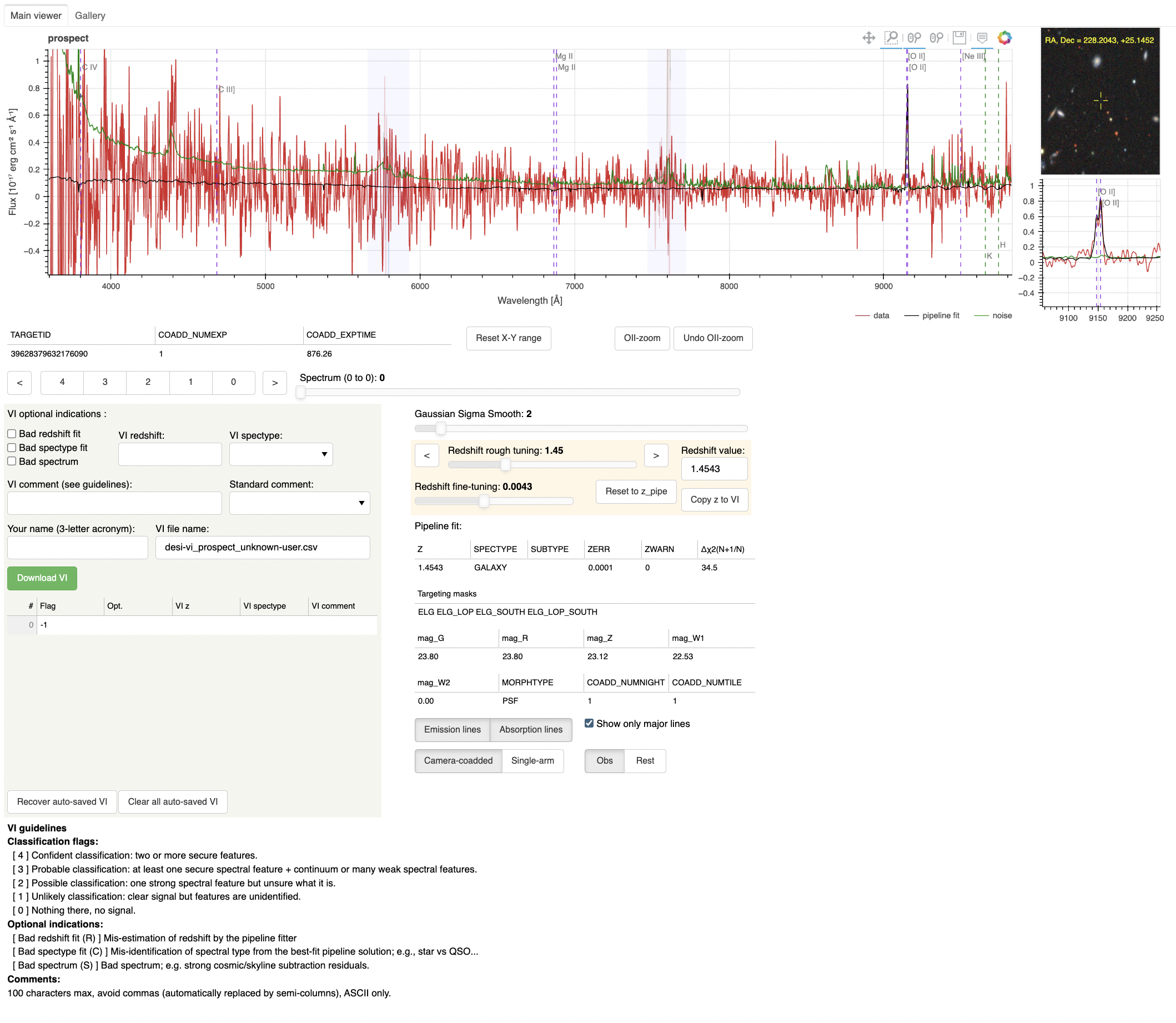}
\caption{A snapshot of the \emph{Prospect} interface used for visual inspection.}
\label{fig:prospect}
\end{figure*}
{\tam

The visual inspection is a collaborative effort contributed by DESI members. 
Here we summarize the key tasks for performing and coordinating the VI effort. 

\textbf{VI tool} --- A dedicated tool, \emph{Prospect}\footnote{\url{https://github.com/desihub/prospect}}, was developed to produce HTML pages containing the spectra and \emph{Redrock} products to be inspected. Individual HTML pages typically include 50 spectra and are directly accessible by collaboration members from any web browser. 
Figure~\ref{fig:prospect} shows an example of the HTML page. 
The main panel displays an individual spectrum, which can be smoothed by the inspector with a variable width Gaussian kernel. The associated noise at each wavelength, as derived from the processing pipeline, can be overlaid, as well as models matching the best fits from \emph{Redrock}, and sets of emission and/or absorption lines. The associated imaging from the Legacy Survey \citep{Dey2019} is displayed next to the spectrum. Inspectors can view relevant information from \emph{Redrock} catalogs, including the ZWARN flag\footnote{\url{https://github.com/desihub/redrock/blob/master/py/redrock/zwarning.py}}, which indicates whether or not the spectrum has issues found by the pipeline, and the best fit $\chi^{2}$ value with respect to second best fit. They can dynamically adjust the redshift for the displayed model and lines. The bottom left panel of the VI pages allows the inspectors to report relevant information, which is saved into an ASCII file for each set of 50 spectra.

\textbf{Recruiting and training} --- To perform the VI, we recruited volunteers from the DESI collaboration. More than 50 DESI members had contributed to the visual inspection of galaxy spectra. In order to obtain consistent results from each inspector, we prepared materials for training the inspectors, including text documents with instructions and videos of experts performing VI and explaining the decision for each spectrum. Meetings were organized for the training as well. All inspectors were required to go through the materials before conducting the VI. After finishing the materials, they were asked to inspect 100 spectra for a given target type of their interest. Their results were then compared to the results from VI experts. If the inconsistent rates between the two results are lower than $2\%$ for BGS, and $10\%$ for ELGs and LRGs, the volunteers can start to perform VI after they reviewed the inconsistent results and agreed with the results. If the inconsistent rates were not lower than the thresholds, before starting the VI, the inspectors were asked to go through the training materials again and they would need to make sure that they understood the inconsistency. Each spectrum was at least inspected by two DESI members and one of the inspectors was an experienced astronomer who had worked on the specific target type. The results from the two inspectors were merged by the VI coordinators into the final VI results and the spectra with inconsistent results were resolved by the merger as described in the main text.}

{}
\end{document}